\documentclass[fleqn,usenatbib]{mnras}

\usepackage[T1]{fontenc}

\DeclareRobustCommand{\VAN}[3]{#2}
\let\VANthebibliography\thebibliography
\def\thebibliography{\DeclareRobustCommand{\VAN}[3]{##3}\VANthebibliography}

\usepackage[pdftex]{graphicx}	% Including figure files
\usepackage{amsmath}	% Advanced maths commands
\usepackage{amssymb}	% Extra maths symbols
\usepackage{epsfig,graphics,subfigure,psfrag,amsmath,amssymb,amscd}
\usepackage{float}
\usepackage{lineno}
\title[Exploration of the origin of 2020 X-ray outburst in OJ 287]{Exploration of the origin of 2020 X-ray outburst in OJ 287}

\author[Huang et al.]{Shifeng Huang$^{1}$, Shaoming Hu$^{1}$,
Hongxing Yin$^{1}$\thanks{E-mail: yinhx@sdu.edu.cn},
Xu Chen$^{1}$, Sofya Alexeeva$^{2}$, and Yunguo Jiang$^{1}$
\\
\\
$^{1}$Shandong Key Laboratory of Optical Astronomy and Solar-Terrestrial Environment,
 School of Space Science and Physics, \\
  ~~~Institute of Space Sciences, Shandong University,
 Weihai, Shandong, 264209, China \\
$^{2}$CAS Key Laboratory of Optical Astronomy, National Astronomical Observatories, Chinese Academy of Sciences, Beijing, 100102, China \\
}

\begin{document}
%\linenumbers

\maketitle

\begin{abstract}
Research into OJ 287 has been ongoing for many years. In 2020 April–June, this source underwent the second highest X-ray outburst (second only to the 2016--2017 outburst) and the mechanism of this outburst is still under debate. In this paper, we discuss two scenarios to explore the origin of the outburst: an after-effect of a black hole–disc impact and a tidal disruption event (TDE). We present the weak correlations of the spectral index versus X-ray flux and the hardness ratio (HR) versus the soft X-ray flux during the outburst, and these features are different from the case in the quiescent state. The correlations are compared with those of the 2016–2017 outburst with the highest X-ray flux in monitoring history. Analysis of the outbursts in 2016–2017 and 2020 shows that the expected time of the X-ray outburst, based on the theory of the after-effect of the black hole–disc impact and the estimation of available data, is inconsistent with historical observations. The soft X-ray spectra, the barely temporal evolution of colour, and the evolution of the HR mean that the 2020 outburst shares similar features with the 2016–2017 outburst, which was considered as a possible candidate for a TDE. Additionally, we find that the predictions of full TDEs ($t^{-5/3}$) and partial TDEs ($t^{-9/4}$) for the soft X-ray decay light curve are well fitted. Our analysis suggests that the 2020 outburst in OJ 287 is probably related to the TDE candidate.
\end{abstract}


\section{Introduction}

Blazars commonly present dramatic variability in light curves and high polarization, and they are the class of active galactic nuclei (AGNs) where the jet is pointing toward the Earth \citep{urry1995}. The variability can be induced by the activity of the jet, the environment variation in the accretion disc and the geometry of the black hole system. Multiwavelength observations can be used as a tool to study the physical mechanisms in AGNs.

Famous for its quasi-periodic optical variability of about 12 yr, OJ 287 is a blazar and the observations of this object have been going on for over 100 yr \citep{dey2018}. Currently, this object is widely considered to host a supermassive black hole binary (SMBHB) system \citep{sillanpaa1988}. The double-peak structure of the optical outburst with 12-yr quasi-periodicity has attracted scientists for a long time. In order to uncover the phenomenon, several scenarios have been proposed, such as disc precession \citep{katz1997}, the beaming effect of the two jets \citep{villata1998}, the precession of the single jet \citep{abraham2000}, and periodic leakage of the gas stream from the circumbinary disc into the SMBHB \citep{tanaka2013}. However, these models are not satisfactory to explain both the extremely low polarization during the quasi-periodic optical outburst and the timing of the events. The most exciting scenario is a black hole–disc impact model. According to this model, in OJ 287, the primary black hole is embedded in the disc, and the orbiting secondary black hole impacts the disc twice in an orbital period \citep{lehto1996}. As a result, the quasi-periodic optical outbursts with double-peak structure in light curves can be observed. So far, this model has been very successful to explain several optical outbursts \citep{valtonen2008,valtonen2016,laine2020}. Notably, the outburst in the summer of 2019 was predicted accurately by \citet{dey2018}, with an uncertainty of only $\sim$4 h. OJ 287 provides a natural laboratory to test general relativity though the quasi-periodic outburst.

In addition to quasi-periodic optical outbursts, there are other more interesting outbursts in OJ 287. During 2016-–2017, a sudden outburst covering from X-ray to radio was observed \citep{komossa2017,kapanadze2018,kushwaha2018,myserlis2018}. Interestingly, throughout the outburst interval, the $\gamma$-rays did not manifest significant activity. Besides, the X-rays exhibited a ‘softer-when-brighter’ tendency in the 2016–2017 outburst \citep{komossa2017,komossa2020,komossa2021a}. One of the black hole–disc impact events occurred in 2013, and its after-effect is considered to be associated with the outburst in 2016-–2017. In this outburst, a perturbation is produced in the impacted point and propagates along with the disc to the jet base, causing its activity \citep{valtonen2017,kapanadze2018,kushwaha2018,komossa2020,komossa2021a,komossa2021b,prince2021b,prince2021a}.

For a star captured by a black hole, if the stellar self-gravity cannot overwhelm the tidal force from the black hole, it will be torn into debris; this process is the so-called tidal disruption event \citep[TDE;][]{hills1975,rees1988}. In most TDE candidates, some phenomena are observed, such as light-curve decay following $\sim t^{-5/3}$ \citep{rees1988}, a blue continuum with broad emission lines in the optical spectra \citep{velzen2020}, and soft X-ray spectra \citep{saxton2020}. Although most TDE candidates have been found in non-active galaxies, theoretical studies suggest that the TDE rate in AGNs is much higher \citep{kennedy2016}. Currently, some TDE candidates have been found, for example, in the following AGNs: SDSS J165726.81+234528.1 \citep{yang2019}, SDSS J022700.77-042020.6 \citep{liu2020}, and Gaia16aax \citep{cannizzaro2020}. \citet{huang2021} have suggested that the 2016--2017 outburst may be related to a TDE in the supermassive black hole system, because of the soft X-ray domination, no significant temporal evolution of hardness ratio (HR), the strengthening of the emission lines after the outburst, and it is well fitted by the $t^{-5/3}$ of the light curves.

In 2020 April--June, a new outburst covering X-ray to radio bands was observed \citep{komossa2020}, exhibiting a ‘softer-when-brighter’ tendency in X-ray, and ‘bluer-when-brighter’ behaviour in optical bands \citep{prince2021b}. Additionally, the spectral index temporal evolution showed a different pattern for the X-ray and ultraviolet (UV)/optical bands \citep{kushwaha2020}. Interestingly, when the X-ray flux reached the maximum value, an Fe absorption line was reported in the spectrum, suggesting the presence of an outflow with a velocity $\sim 0.1c$ \citep{komossa2020}. Similar to the case in 2016--2017, the $\gamma$-rays did not show any significant activity during the outburst in 2020 April--June. During the outburst, the spectral energy distributions (SEDs) manifested a component similar to the high-energy peak \citep{kushwaha2020,prince2021b,singh2021}, just like the case in the 2016--2017 outburst \citep{kushwaha2018}. Recently, the 2020 outburst has been studied by some researchers, and the after-effect of black hole–disc impact is thought to be a probable origin \citep{komossa2020,komossa2021a,kushwaha2020,komossa2021b,prince2021b,prince2021a}.

In this work, we investigated the outburst in 2020 April–June, through the multiwavelength light curves, the evolution of the X-ray spectra, the HR and UV/optical colours. In order to study the physical mechanism of the outburst, we compared properties of the outbursts in both 2016–2017 and 2020. Given that the after-effect of black hole–disc impact is previously thought to be a possible origin, based on this scenario, we estimated the peak time intervals of the X-rays. Additionally, we proposed that a TDE can be considered as the origin, and so the soft X-ray light curve and stellar property were analysed. In Section 2, we introduce the data reductions. In Section 3, we present the multiwavelength light curves, the evolution of the HR and spectral index, and UV/optical colour variations. Then, in Section 4, we discuss the origin of the outburst in 2020 though two scenarios: the after-effect of the black hole–disc impact and the TDE. Finally, we discuss the results and conclude in Section 5. The cosmological parameters from the Wilkinson Microwave Anisotropy Probe ninth-year results (WMAP9; Hinshaw et al. 2013) are used in this work: $H_0=69.3~\text{km}~\text{s}^{-1}\text{Mpc}^{-1}$, $\Omega_M=0.287$, and $\Omega_{\Lambda}=0.713$.

 In this work, we investigated the outburst in 2020 April--June, through the multiwavelength light curves, evolution of the X-ray spectra, HR and UV/optical colours. In order to study the physical mechanism of the outburst, we compared properties of the outbursts in both 2016--2017 and 2020. Given that the aftereffect of black hole-disk impaction is previously thought to be a possible origin, based on this scenario, we estimated the peak time intervals of the X-rays. Additionally, we proposed that a TDE can be considered as the origin, then the soft X-ray light curve and stellar property were analyzed. In section~\ref{sec:observation}, the data reductions are introduced, and in section~\ref{sec:results}, the multiwavelength light curves, evolution of HR and spectral index, and UV/optical colours variations are presented. Then, in section~\ref{sec:discussion}, we discuss the origin of the outburst in 2020 though two scenarios, the aftereffect of the black hole-disk impaction and TDE. Finally, we conclude the results in section~\ref{sec:conclusion}. The cosmological parameters from the Wilkinson Microwave Anisotropy Probe's ninth-year results (WMAP9) \citep{hinshaw2013} are used in this work: $H_0=69.3~\text{km}~\text{s}^{-1}\text{Mpc}^{-1}$, $\Omega_M=0.287$, and $\Omega_{\Lambda}=0.713$.

\section{Observation and Data Reduction}\label{sec:observation}
\subsection{Fermi $\gamma$-ray data}
The high-energy $\gamma$-ray telescope Large Area Telescope (\emph{Fermi}-LAT), was equipped on the Fermi Gamma-ray Space Telescope mission, monitoring the energy range from 20 MeV to 300 GeV \citep{atwood2009}. The $\gamma$-ray data of the object were downloaded from the \emph{Fermi}-LAT data website\footnote{\url{https://fermi.gsfc.nasa.gov/cgi-bin/ssc/LAT/LATDataQuery.cgi}} with the interval from 2008 August 1 to 2021 March 1.

We analyzed the data with standard unbinned likelihood tutorials\footnote{\url{https://fermi.gsfc.nasa.gov/ssc/data/analysis/scitools/likelihood_tutorial.html}} through the Fermi Science Tools \texttt{Fermitools} version 1.0.10. With the radius of the search region of $10^{\circ}$, we ran \texttt{gtselect} to select the source region, centering on the coordinates of the object, and the event data were extracted. Then the good time intervals were set with \texttt{gtmktime}. To create the counts map of the degree region of interest (ROI), \texttt{gtbin} was run, afterwards, we obtained the exposure map through the tasks \texttt{gtltcube} and \texttt{gtexpmap}. The current Galactic diffuse emission model \rm{gll\_iem\_v07.fits} and the corresponding model for the extragalactic isotropic diffuse emission iso\_P8R3\_SOURCE\_V2\_v1.txt, were used as the background models to create the XML files with \rm{make4FGLxml.py}. \texttt{gtdiffrsp} was executed while creating the diffuse source responses. Finally, we obtained the flux of the object with \texttt{gtlike}. In this work, the threshold value of the test statistics (TS) is set to be 5.

\subsection{Swift XRT and UVOT observations}
\subsubsection{X-ray data}
The \emph{Swift} X-Ray Telescope (XRT), covering from 0.3 to 10 keV, is an X-ray detector on the Neil Gehrels \emph{Swift} observatory \citep{burrows2005}. The data from 2005 May 20 to 2021 April 19 were retrieved from the High Energy Astrophysics Science Archive Research Center (HEASARC) website\footnote{\url{https://heasarc.gsfc.nasa.gov/db-perl/W3Browse/w3browse.pl}} and all the data were analyzed from level I.

The \emph{Swift} data were reduced with \texttt{HEASoft 6.28}. From the raw data, we ran \texttt{xrtpipeline} to obtain the level II data including the clean event files. Centering on the object, we extracted a source region file using a circle with radius of 47 arcsec and the background region file was extracted by an annulus with an inner radius of 165 arcsec and outer radius of 235 arcsec. The pile-up effect was checked when the counts rate $> 0.5 ~\text{count}~\text{s}^{-1}$ for photon counting (PC) mode and above $> 100~\text{count}~\text{s}^{-1}$ for windowed timing (WT) mode\footnote{\url{https://www.swift.ac.uk/analysis/xrt/pileup.php}}. We extracted the source region through the annulus with inner radius of 4 arcsec and outer radius of 47 arcse. The light curves, spectra and ancillary response files were generated by \texttt{xrtproducts} and the net light curves were generated with \texttt{lcmath}.

The spectra were grouped by \texttt{grppha} with a minimum of 20 photons per bin and 10 photons per bin for the spectra of WT mode and PC mode, respectively. The response matrix file swxwt0to2s6\_20131212v015.rmf for WT mode and swxpc0to12s6\_20130101v014.rmf for PC mode were used in the analysis.

\subsubsection{UV-optical data}
The UV/Optical Telescope (UVOT) is one of the instruments on \emph{Swift} observatory, with the effective wavelength range 170--600 nm,  and seven filters (\textsl{v}, \textsl{b}, \textsl{u}, \textsl{uvw1}, \textsl{uvm2}, \textsl{uvw2}, and white) were equipped \citep{roming2005}. Following the recommended threads\footnote{\url{https://www.swift.ac.uk/analysis/uvot/index.php}}, we ran \texttt{uvotsource} in \texttt{HEASoft 6.28} to generate the light curves. For the \textsl{v}, \textsl{b}, and \textsl{u} bands, the source region files were extracted by the circle with radius 10 arcsec, centering on the object and nearing the source. The background region was extracted using a circle with a radius of 50 arcsec. For \textsl{uvw1}, \textsl{uvw2}, and \textsl{uvm2} bands, the source and background region files were extracted by the circle with the radius of 15 and 75 arcsec, respectively. In this work, we use the Galactic extinction of 0.11, 0.15, 0.18,
0.24, 0.34, and 0.32 mag in the \textsl{v}, \textsl{b}, \textsl{u}, \textsl{uvw1}, \textsl{uvm2}, and \textsl{uvw2} bands, respectively \citep{kapanadze2018}.

\section{Results}\label{sec:results}
\subsection{Multiwavelength light curves of OJ 287}

\begin{figure}
  \centering
  \includegraphics[width=0.45\textwidth]{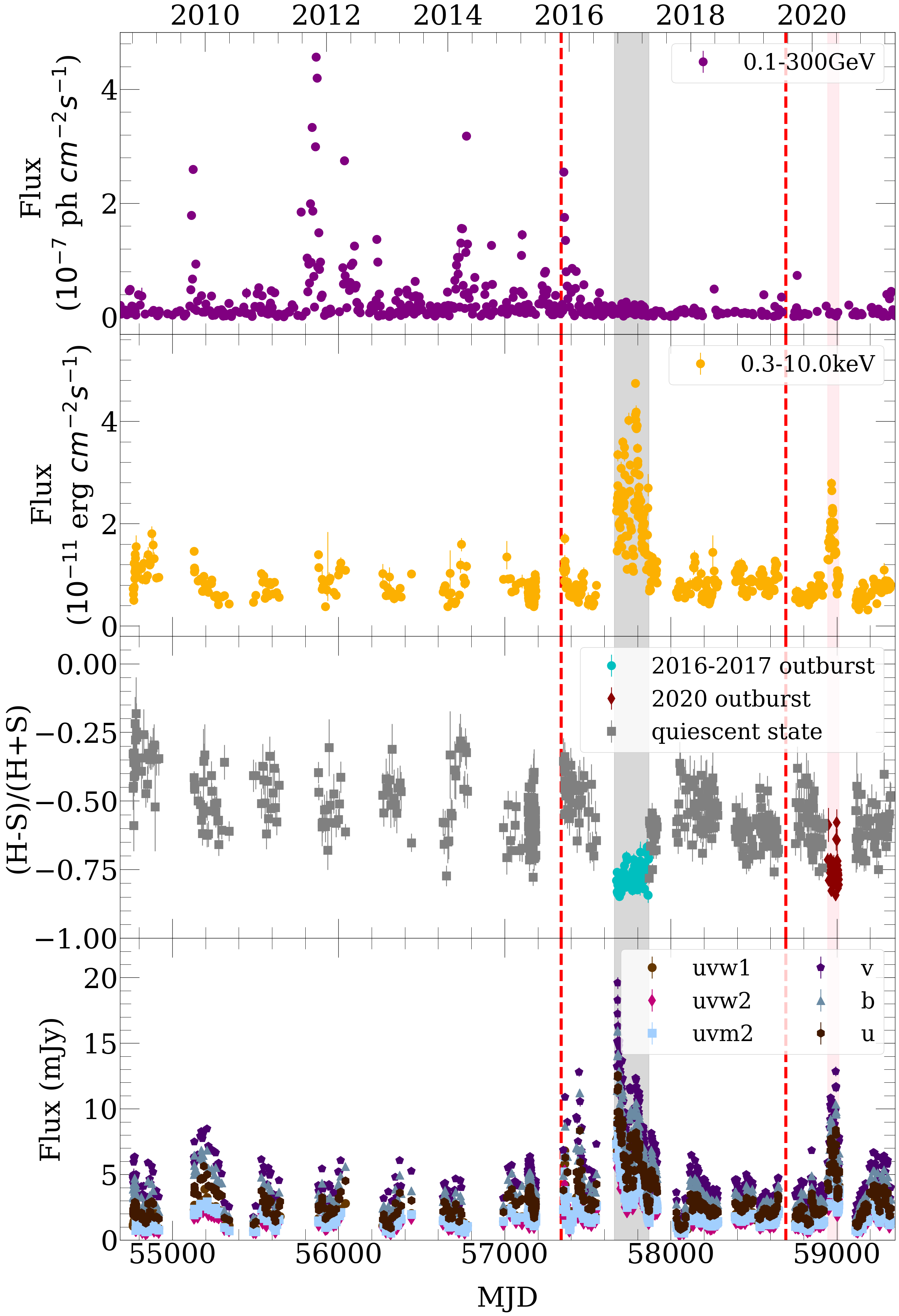}
  \caption{The multiwavelength light curves during 2008 August–2021 May. In the top panel, the 5-d binned $\gamma$-ray light curve from 0.1 to 300 GeV is shown. The second panel plots the X-ray (0.3–10.0 keV) light curve. The third panel shows the HR for X-ray, and the two outbursts are marked by two different colours. The bottom panel shows the light curves of the UV/optical bands. The grey and pink shading in each panel marks the intervals of the 2016–2017 and 2020 X-ray outbursts, respectively. The red dashed lines in each panel mark the starting time of the optical outburst caused by black hole–disc impact \citep{dey2018}. We should note that there are 620 UV/optical data points for each band in the bottom panel, and because of the lack of high-cadence observations, some details in the 2015 optical outburst are missed. }\label{fig:mw_lc}
\end{figure}

The multiwavelength light curves from $\gamma$-ray to optical bands obtained from 2008 August to 2021 May are shown in Figure~\ref{fig:mw_lc}.
The 2016--2017 X-ray outburst was for $\sim$ 200 days,
whereas the 2020 outburst was only two months. The 2016–2017 X-ray outburst is the brightest in monitoring history and the 2020 outburst is the second highest.
The outbursts covered from X-ray to radio bands \citep{komossa2017,kapanadze2018,kushwaha2018,komossa2020,komossa2021a,komossa2021b,prince2021a}. It should be noted that, during these two outbursts, no prominent $\gamma$-ray activity was detected. Before the 2016-2017 X-ray outburst, \emph{Fermi-LAT} had detected multiple outbursts in $\gamma$-rays, while at the same time observations by \emph{Swift} showed that the X-rays were still in a quiescent state during this period.
The X-ray outburst in 2016--2017, first reported by \cite{komossa2017},  exhibited the `softer-when-brighter' behavior and it reached its brightest at the X-ray monitoring \citep{kapanadze2018,kushwaha2018,komossa2020,komossa2021a,komossa2021b,prince2021b,huang2021}.
The `softer-when-brighter' behavior also can be seen in the X-ray outburst in 2020. The HR is defined by $\frac{H-S}{H+S}$, where H and S denote the counts rate of the hard X-ray (2.0--10.0 keV) and the soft X-ray (0.3--2.0 keV), respectively. During the X-ray outburst, the mean HR is $-0.785\pm 0.002$ for the 2016--2017 outburst and $-0.754\pm0.004$ for the 2020 outburst. For the quiescent state, the mean HR is $-0.535\pm0.032$.

\subsection{The outburst in 2020}
\subsubsection{Spectral fitting of the X-ray}
We fitted the X-ray spectra with the absorbed redshift power-law model (\texttt{tbabs*zpowerlw}) with the hydrogen column of $N_{H}=2.49\times 10^{20} \text{cm}^{-2}$ \citep{kapanadze2018}. The evolution of the spectral index $\Gamma$ can be seen in the Figure~\ref{fig:2020burst}. The mean value of the spectral index during the outburst is $2.510\pm{0.026}$ and $1.898\pm{0.011}$ for the quiescent state.

\subsubsection{The light curves of the outburst}
\begin{figure}
  \centering
  \includegraphics[width=0.45\textwidth]{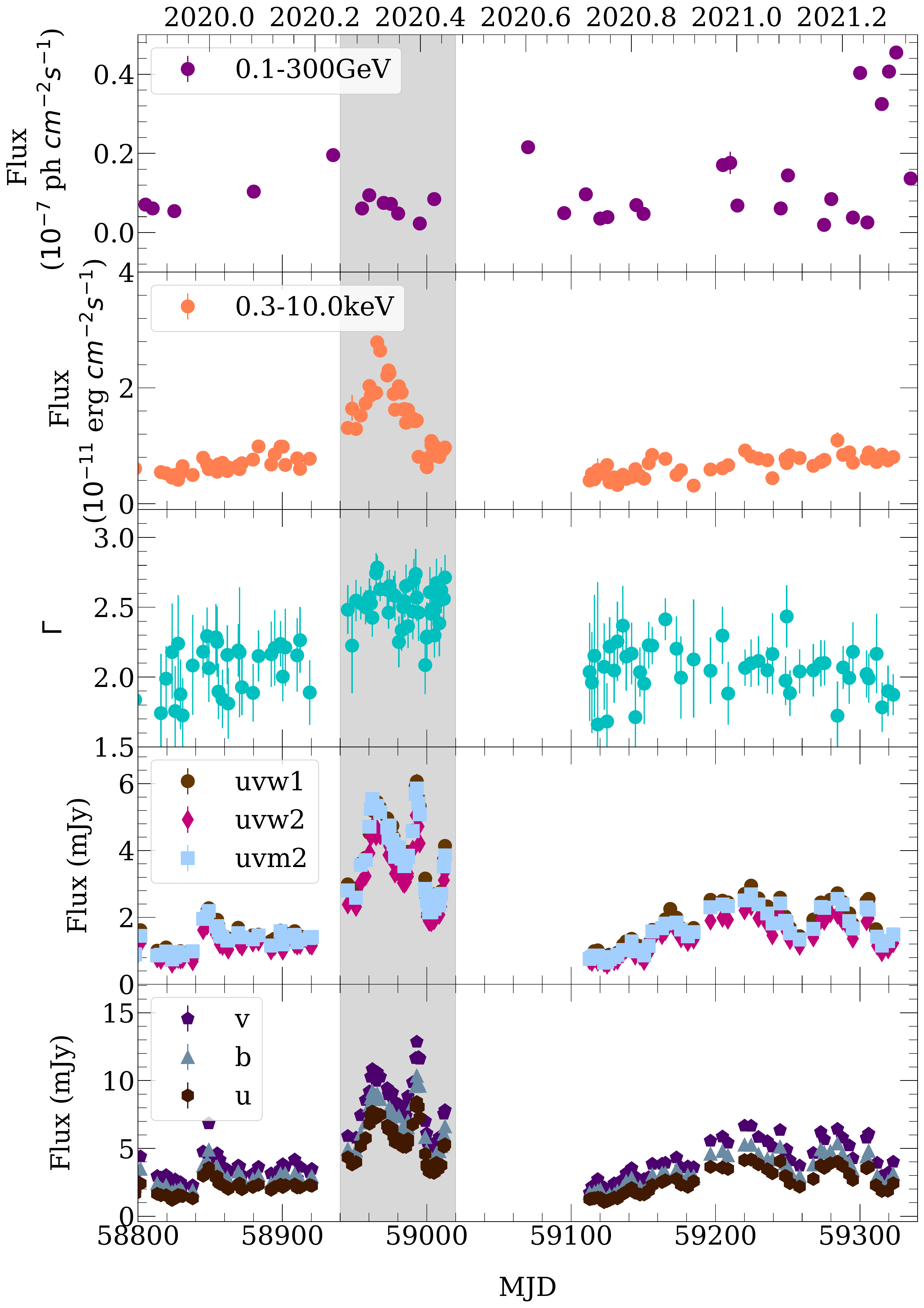}
  \caption{The outburst of OJ 287 in 2020. The top panel shows the 0.1--300 GeV variability of the $\gamma$-ray with 5-d bins. In the second panel, the X-ray light curve from 0.3 to 10.0 keV is plotted. The temporal evolution of the spectral index can be seen in the middle panel. In the bottom two panels, the light curves of UV and optical bands are presented. The grey shaded regions in each panel mark the outburst in 2020. The $\gamma$-ray light curve is integrated from 100 MeV to 300 GeV, so the error bars in the top panel are smaller than the data points. } \label{fig:2020burst}
\end{figure}

The multiwavelength light curves during 2019 November--2021 May are shown in Figure~\ref{fig:2020burst}. The $\gamma$-rays did not show prominent activity at that time. During the outburst, the mean flux of the X-rays (0.3--10.0 keV) is $1.465\pm{0.013}\times 10^{-11}\text{erg}~\text{cm}^{-2}~\text{s}^{-1}$, while the mean flux for the quiescent state
is $0.750\pm{0.004}\times 10^{-11}\text{erg}~\text{cm}^{-2}~\text{s}^{-1}$. Around MJD 58962, the peak of the X-ray flux was observed \citep{komossa2020}. After this, the X-rays continued to decay, while the UV/optical had rebrightening at around MJD 58993. Besides, the minimum value of the UV/optical flux in the outburst was observed around MJD 59001, and after which the light curves exhibited a rising tendency. However, because of the lack of ongoing observations, the third potential peak cannot be determined.

\subsubsection{The rebrightening of UV/optical bands during the outburst} 
\begin{figure}
  \centering
  \includegraphics[width=0.45\textwidth]{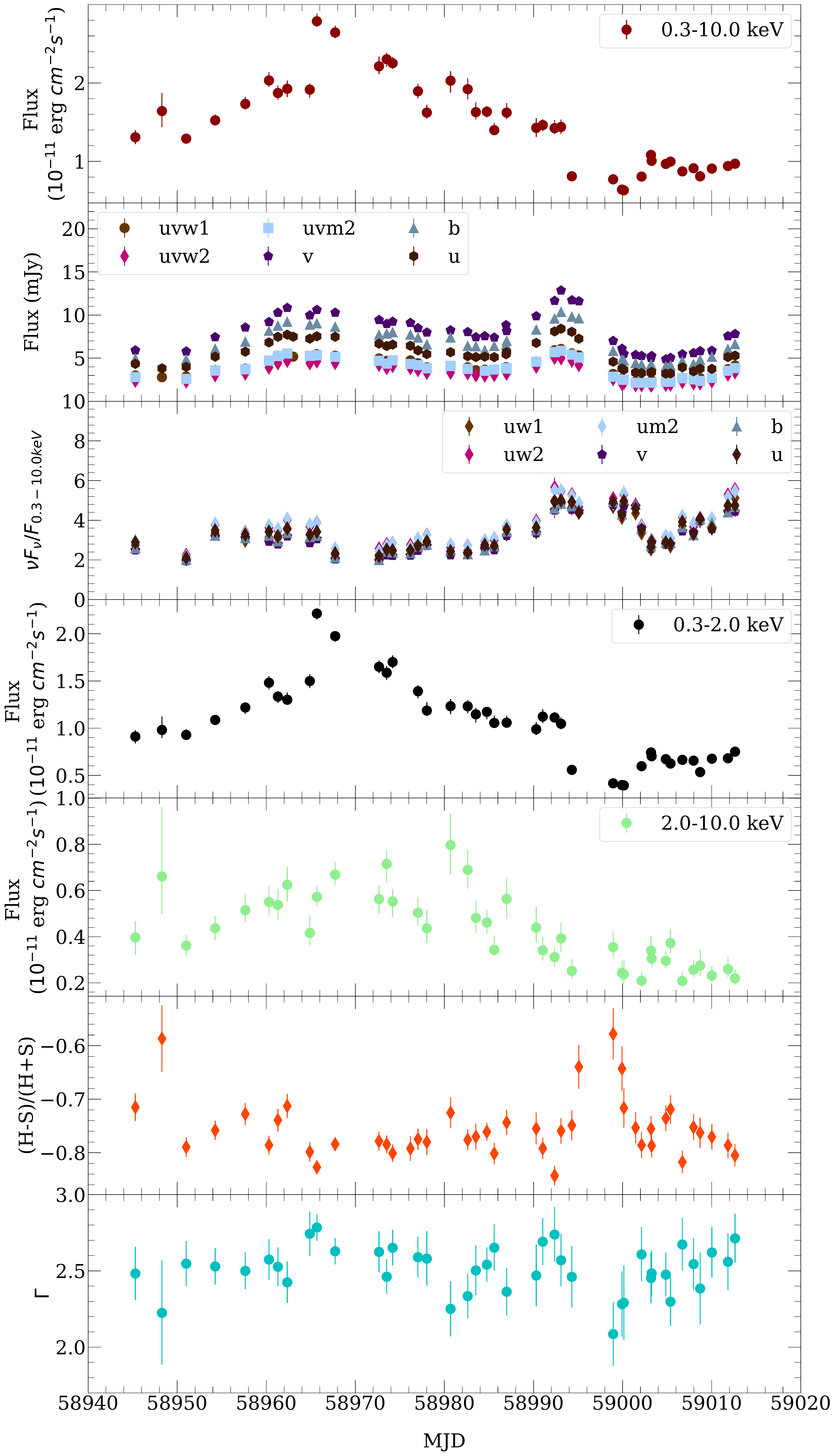}
  \caption{ The evolution of the spectral index and the HR with the X-ray flux. The left panel plots the spectral index as a function of the X-ray flux in 0.3--10.0 keV and the right panel shows the evolution of the HR with the soft X-ray flux. We separately fitted the data points for three epochs (2016--2017 outburst, 2020 outburst, and quiescent state) with linear fitting. The data points of the 2016--2017 outburst, the 2020 outburst, and the quiescent state are shown by the dashed, solid, and dot-dashed lines, respectively.}\label{fig:uv_index}
\end{figure}

Around MJD 58993, mainly in UV/optical bands, the second peak in the outburst was observed. The details of the second flare can be seen in Figure~\ref{fig:uv_index}. The X-ray light curve performed a slight enhancement, while the UV/optical bands manifested much more dramatically. The ratio of UV/optical flux to the X-ray evolved smoothly during the first flare. In the second flare, the ratio increased, where around MJD 58992, the value of the uvw2 flux ratio to the X-ray reached a maximum of $5.662\pm{0.485}$. During the second flare, in MJD 58999, the HR exhibited a high level and the value reached up to $-0.578\pm{0.048}$. Meanwhile, the X-ray spectral index dropped to $2.086_{-0.209}^{+0.210}$, the minimum value during the outburst.

\subsection{The correlation analysis}
\begin{figure}
  \centering
  \includegraphics[width=0.45\textwidth]{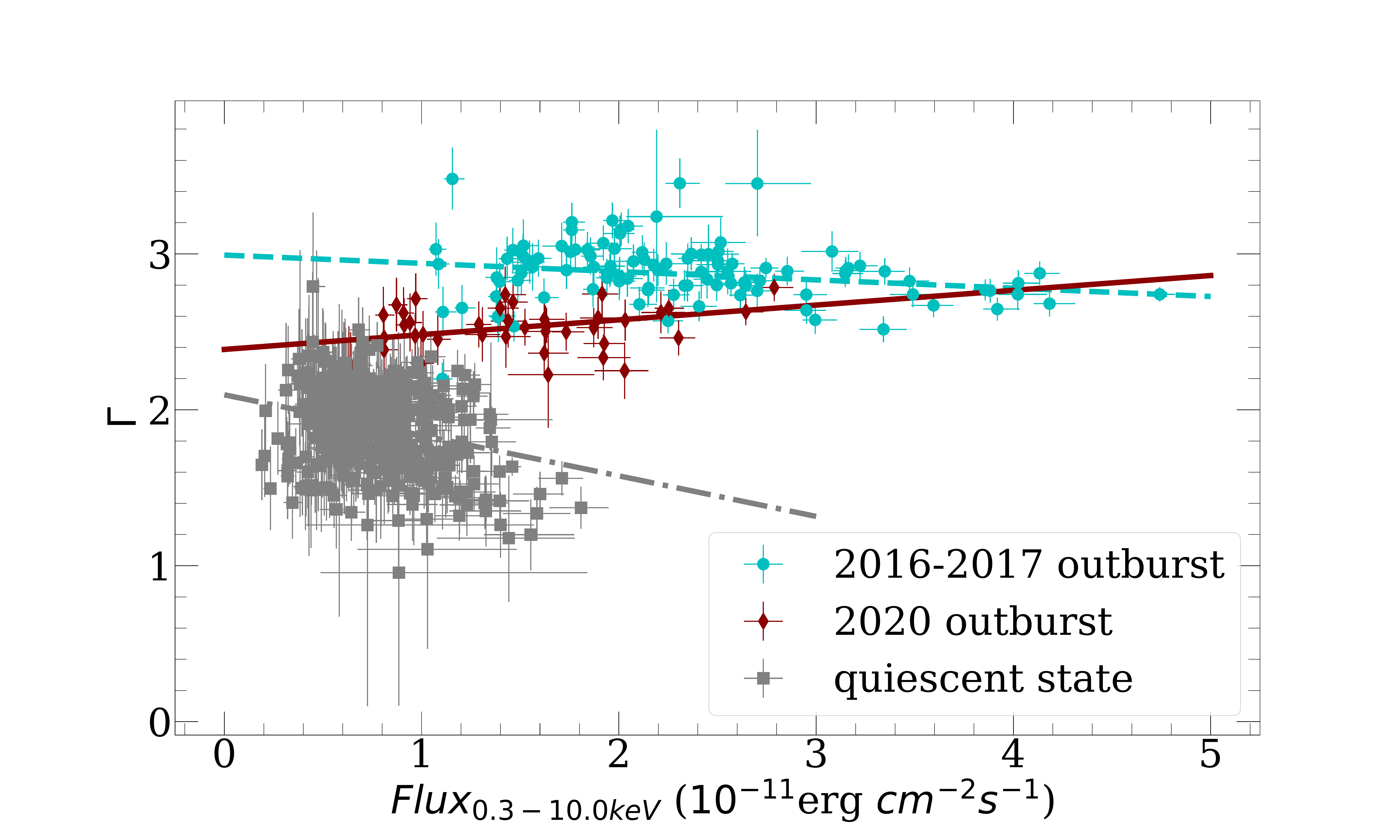}
  \includegraphics[width=0.45\textwidth]{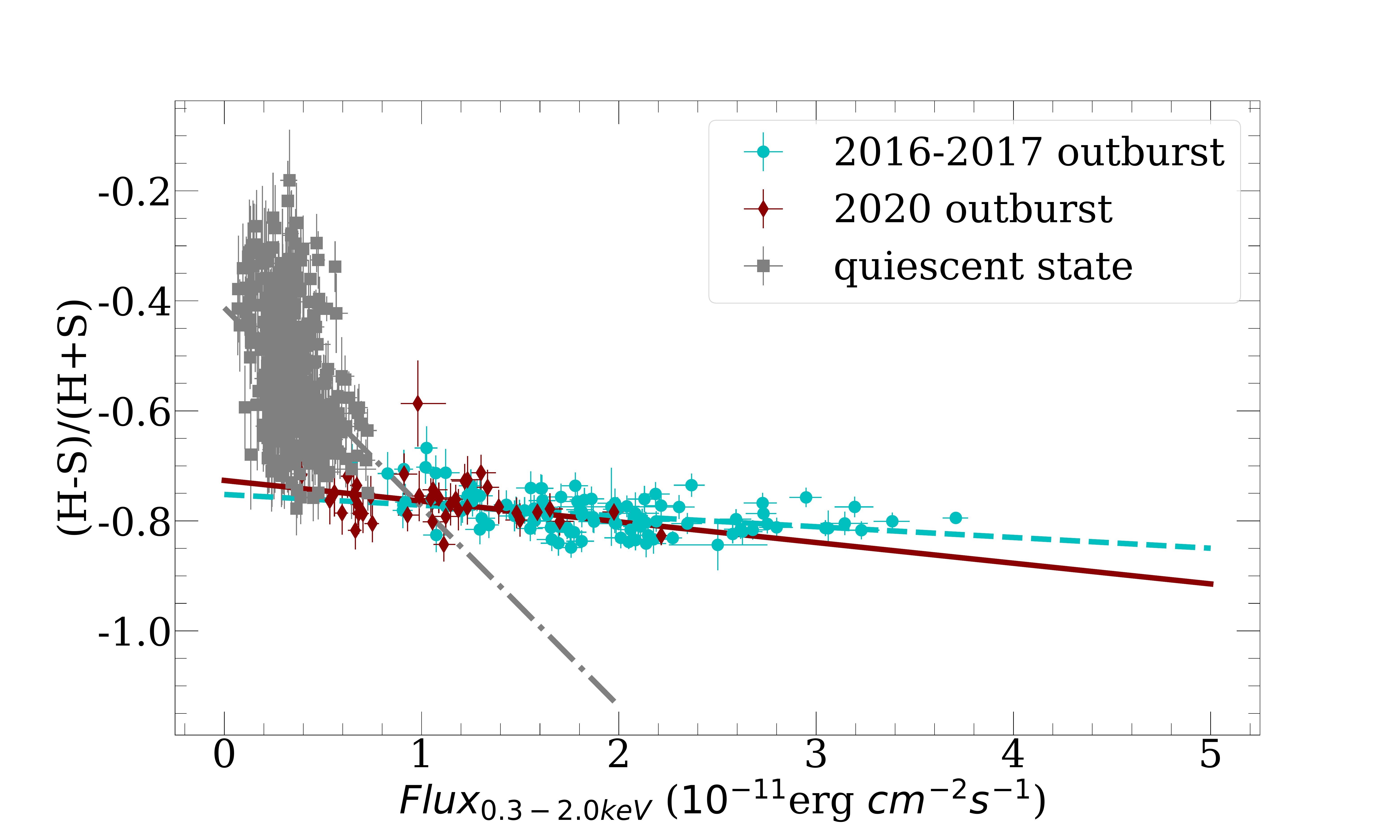}
  \caption{The evolution of spectral index and HR with the X-ray flux. The left panel plots the spectral index as a function of the X-ray flux in 0.3--10.0 keV and the right panel shows the evolution of HR with the soft X-ray flux. We separately fitted the data points for three epochs (2016--2017 outburst, 2020 outburst and quiescent state) with the linear fitting. The data points of 2016--2017 outburst, 2020 outburst and the quiescent state were fitted by the dash, solid line, and dot dash lines, respectively. }\label{fig:index_flux}
\end{figure}

The spectral index and HR are shown as a function of the X-ray flux in Figure~\ref{fig:index_flux}. The correlation between spectral index and flux, HR and soft flux were adopted by Spearman-rank test. For three epoches, we fitted the data points separately through the python package \emph{emcee} \citep{foreman2013} with 50 walkers and 10000 iterations.

\begin{table*}
\centering
\caption{Correlation analysis of spectral index versus flux and HR versus soft X-ray flux. Here, $r_s$ is the Spearman-rank test correlation coefficient and p denotes p-value. The slope and intercept are obtained by a linear fit. The uncertainties are generated from the 16th, 50th, and 84th percentiles of the samples in the marginalized posterior distributions.}
\begin{tabular}{lcccccccc}
%\tabletypesize{\scriptsize}
\hline
 & \multicolumn{4}{c}{spectral index versus flux} & \multicolumn{4}{c}{HR versus soft X-ray flux} \\ \hline
 & $r_s$ & p & Slope & Intercept & $r_s$ & p & Slope &
Intercept \\ \hline
2016--2017 outburst  & -0.228 & 0.021 & $-0.053\pm{0.021}$ & $2.992^{+0.054}_{-0.053}$ & -0.439 & $3.427\times 10^{-6}$ & $-0.020\pm{0.005}$ & $-0.752\pm{0.011}$ \\
2020 outburst & 0.235 & 0.139 & $0.095\pm{0.038}$ & $2.387^{+0.068}_{-0.067}$  & -0.341 & 0.029 & $-0.038\pm{0.011}$ & $-0.727\pm{0.014}$ \\
Quiescent state & -0.209 & $3.150\times 10^{-6}$ & $-0.260^{+0.043}_{-0.042}$ & $2.096\pm{0.036}$ &  -0.401 & $2.136 \times 10^{-20}$ & $-0.362^{+0.039}_{-0.040}$ & $-0.413\pm{0.014}$ \\ \hline
\end{tabular}
\label{tab:index_hr_evolution}
\end{table*}

The patterns of both 2016--2017 and 2020 outbursts are different from the pattern of the quiescent state. In the case of the correlation between spectral index and flux, the Spearman-rank correlation coefficients ($r_s$) showed the similar values for both 2016--2017 outburst and quiescent state.
Among them, $r_s=-0.228$ with p-value (p) for null hypothesis is 0.021 ($<0.05$) for 2016--2017 outburst. For the quiescent state, we obtain $r_s=-0.209$ and p-value of $3.150\times 10^{-6}$ ($\ll 0.05$). However, for the 2020 outburst, $r_s=0.235$ and p-value of 0.139. In the linear fitting results, the slope of the data points of the quiescent state is $-0.260^{+0.043}_{-0.042}$ while the case in both 2016--2017 and 2020 outburst are $-0.053\pm{0.021}$ and $0.095\pm{0.038}$, respectively, which exhibited the flattening. More details can be seen in Table~\ref{tab:index_hr_evolution}. In the correlation analysis between HR and soft flux, we obtained $r_s$ values of $-0.439$, $-0.341$ and $-0.401$ for the case of quiescent state, 2016--2017 outburst and 2020 outburst, respectively. The p-values for the three epochs are less than 0.05.

\subsection{UV/optical colour evolution}
\begin{figure*}
  \centering
  \subfigure [2016--2017 outburst] {
  \includegraphics[width=0.45\textwidth]{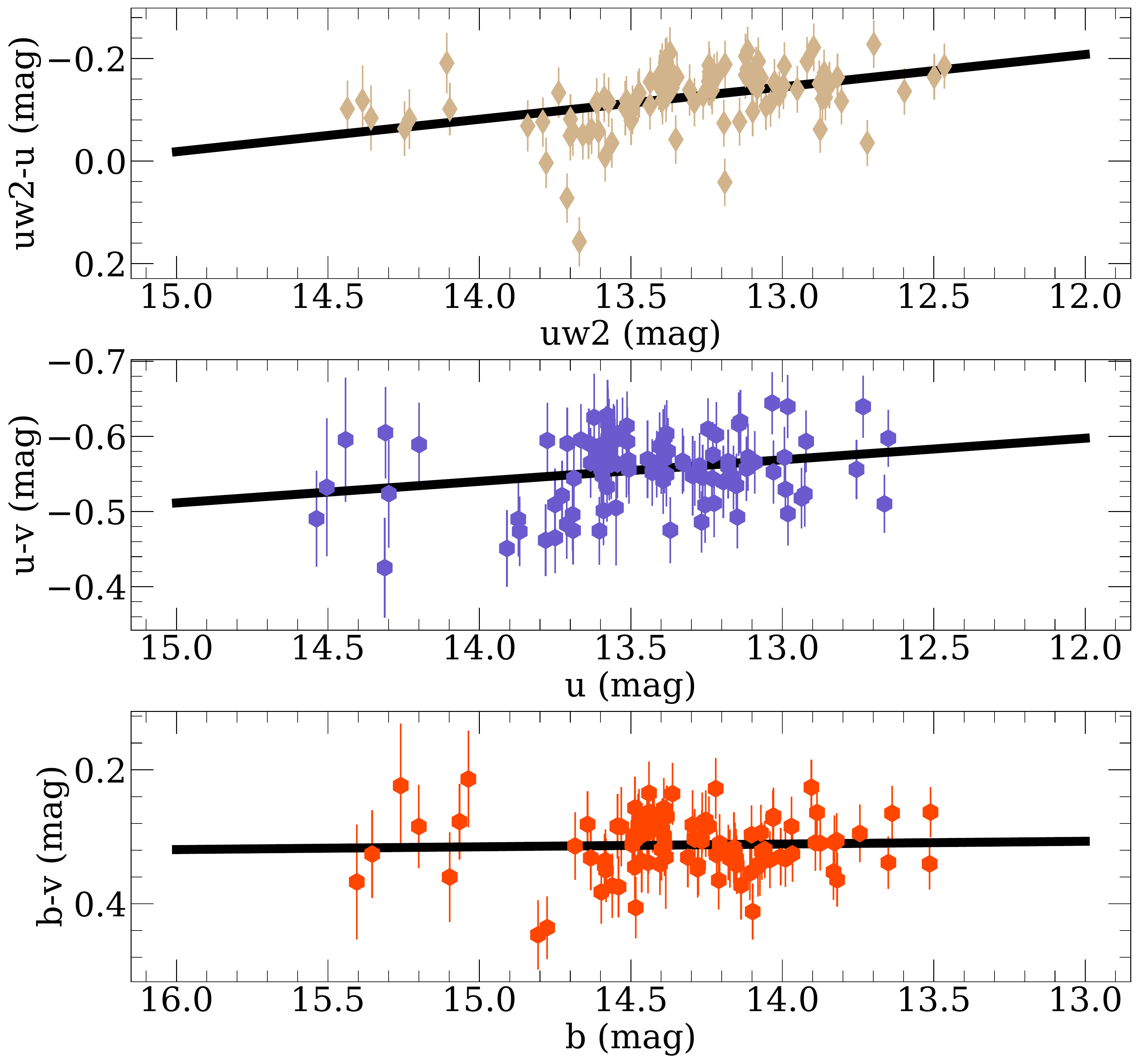}}
  \subfigure [2020 outburst] {
  \includegraphics[width=0.46\textwidth]{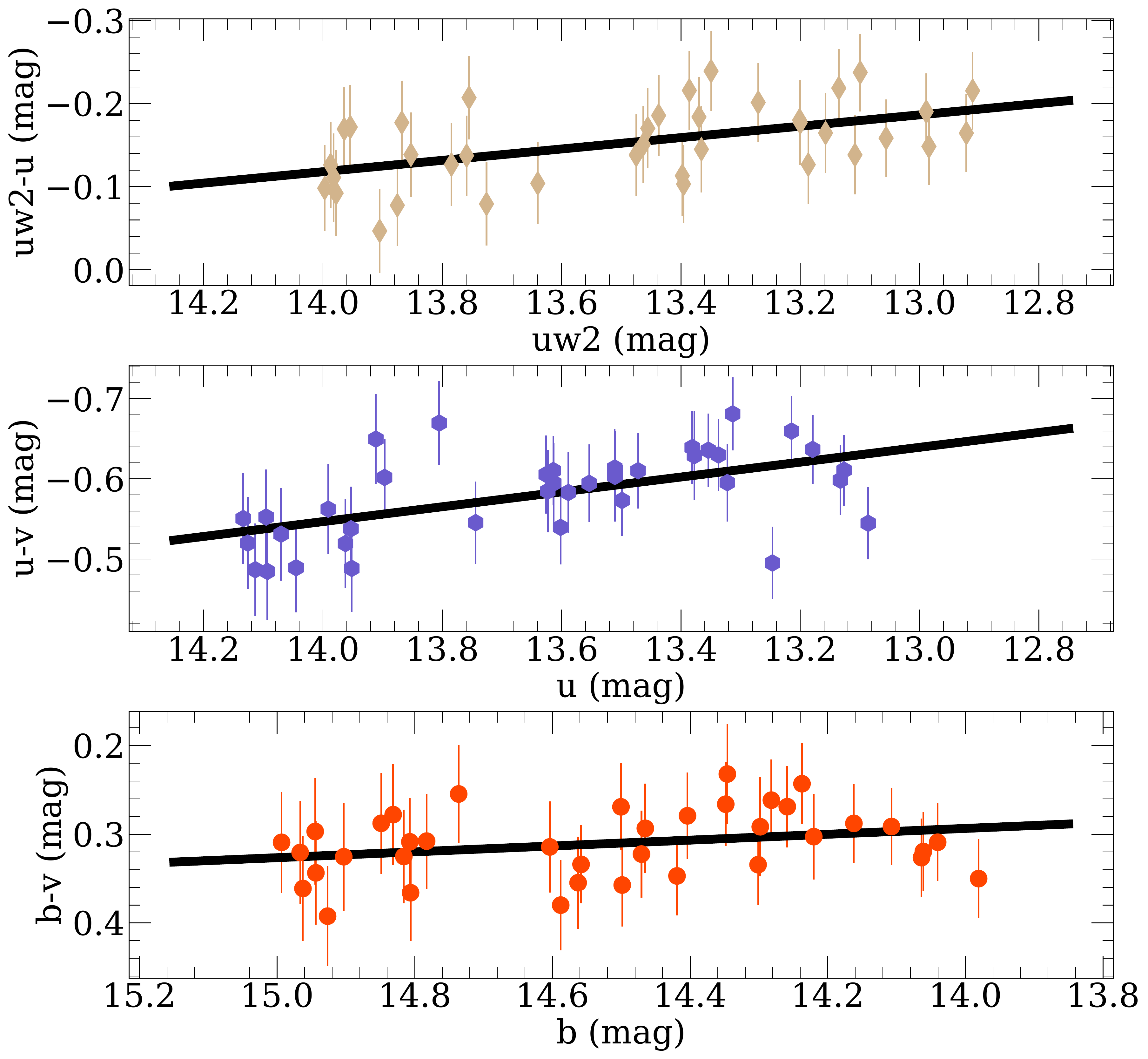}}
  \caption{Colour evolution with time during the 2016--2017 and 2020 outbursts. The black solid line in each panel denotes the linear fitting.}\label{fig:color}
\end{figure*}

We calculated the colour evolution of \textsl{uw2-v}, \textsl{u-v}, and \textsl{b-v}. In Figure~\ref{fig:color}, the colour-magnitude correlation is presented. We noticed that the UV/optical bands in 2020 outburst magnified a `bluer-when-brighter' pattern, which is consistent with the results of \citet{prince2021a}. Additionally, in the case of the 2016--2017 outburst, a similar behaviour can be seen. The correlation and linear fitting results are listed in Table~\ref{tab:color_evolution}. Both \textsl{uw2-v} and \textsl{u-v} exhibit a strong correlation between colour and magnitude, while only b–v shows a weaker correlation in both the 2016–2017 and 2020 outbursts. This tendency is consistent with the results of \citet{gupta2019}.

\begin{table*}
\centering
\caption{Colour evolution for UV/optical bands. Here, $r_s$ is the Spearman rank test correlation coefficient and p denotes the p-value. The slope and intercept are obtained by a linear fit. The uncertainties are generated from the 16th, 50th, and 84th percentiles of the samples in the marginalized posterior distributions. \label{tab:color_evolution}}
\begin{tabular}{lcccccccc}
\hline
 & \multicolumn{4}{c}{2016--2017 outburst} & \multicolumn{4}{c}{2020 outburst} \\ \hline
 & $r_s$ & p & Slope & Intercept & $r_s$ & p & Slope &
Intercept  \\ \hline
uw2 vs uw2-u & 0.480 & $5.549\times 10^{-7}$ & $0.067\pm{0.014}$ & $-1.02^{+0.191}_{-0.186}$ & 0.540 & $2.664\times 10^{-4}$ & $0.069\pm{0.022}$ & $-1.082^{+0.291}_{-0.289}$ \\
u vs u-v & 0.166 & 0.103 & $0.029\pm{0.014}$ & $-0.946^{+0.184}_{-0.185}$ & 0.595 & $4.035\times 10^{-5}$ & $0.096\pm{0.024}$ & $-1.889^{+0.323}_{-0.325}$ \\
b vs b-v & 0.026 & 0.799 & $0.010\pm{0.014}$ & $0.0169\pm{0.199}$ & 0.303 & 0.054 & $0.036^{+0.026}_{-0.025}$ & $-0.216^{+0.365}_{-0.372}$ \\
time vs uw2-u & 0.014 & 0.891 & $-2\pm{11}\times10^{-5}$ & $-0.123\pm{0.012}$ & 0.068 & 0.671 & $1.400^{+4.1}_{-4.0}\times 10^{-4}$ & $-0.163\pm{0.018}$ \\
time vs u-v & -0.073 & 0.478 & $-5\pm{8}\times 10^{-5}$ & $-0.553\pm{0.009}$ & 0.697 & $4.051\times10^{-7}$ & $1.990^{+0.41}_{-0.42}\times 10^{-3}$ & $-0.667\pm{0.017}$ \\
time vs b-v & -0.064 & 0.530 & $-2\pm{8}\times 10^{-5}$ & $0.315\pm{0.009}$ & 0.262 & 0.098 & $5.5^{+4.1}_{-4.2}\times 10^{-4}$ & $0.287\pm{+0.018}$ \\ \hline
\end{tabular}

\end{table*}

\begin{figure*}
  \centering
  \subfigure [2016--2017 outburst] {
  \includegraphics[width=0.45\textwidth]{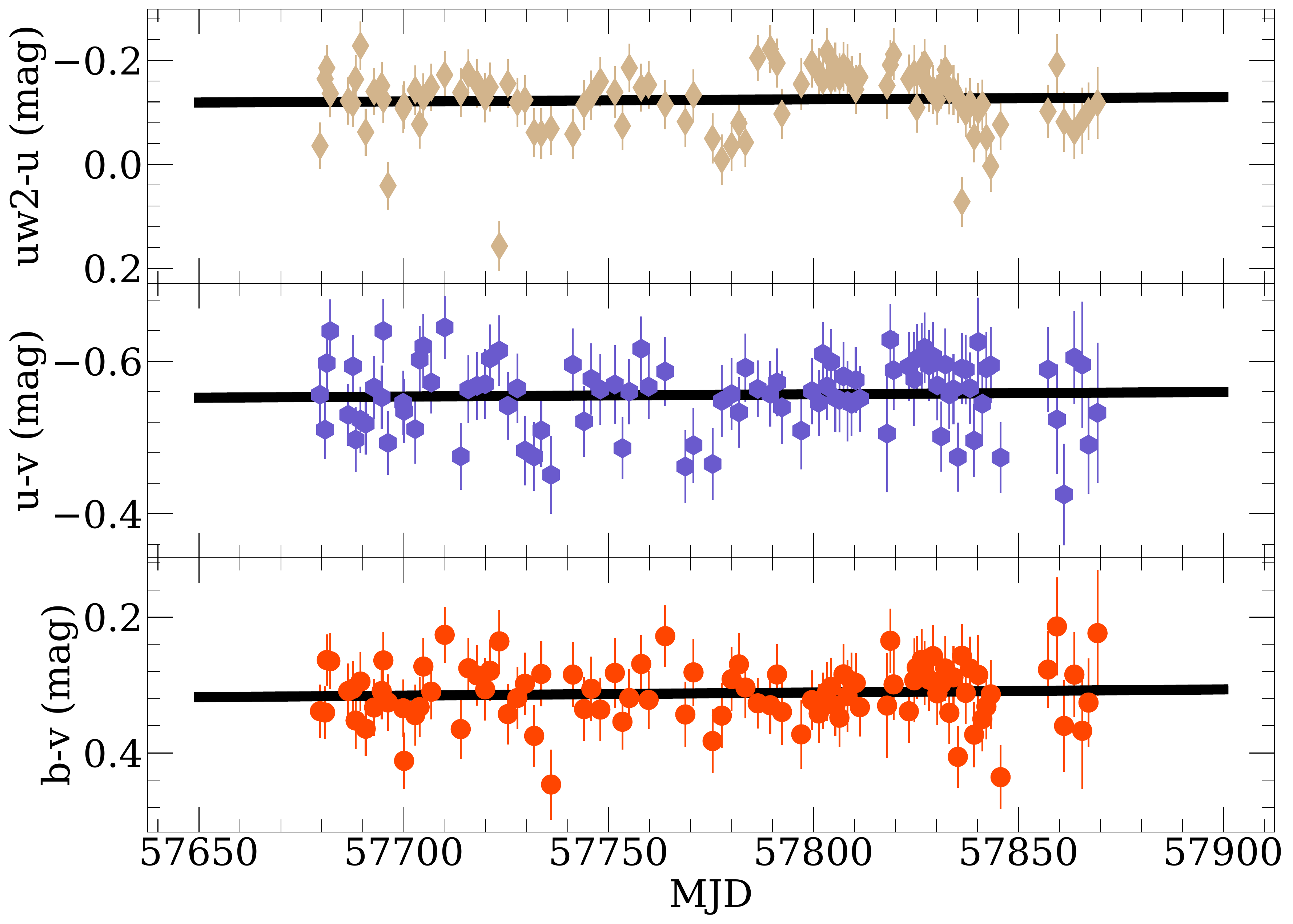}}
  \subfigure [2020 outburst] {
  \includegraphics[width=0.45\textwidth]{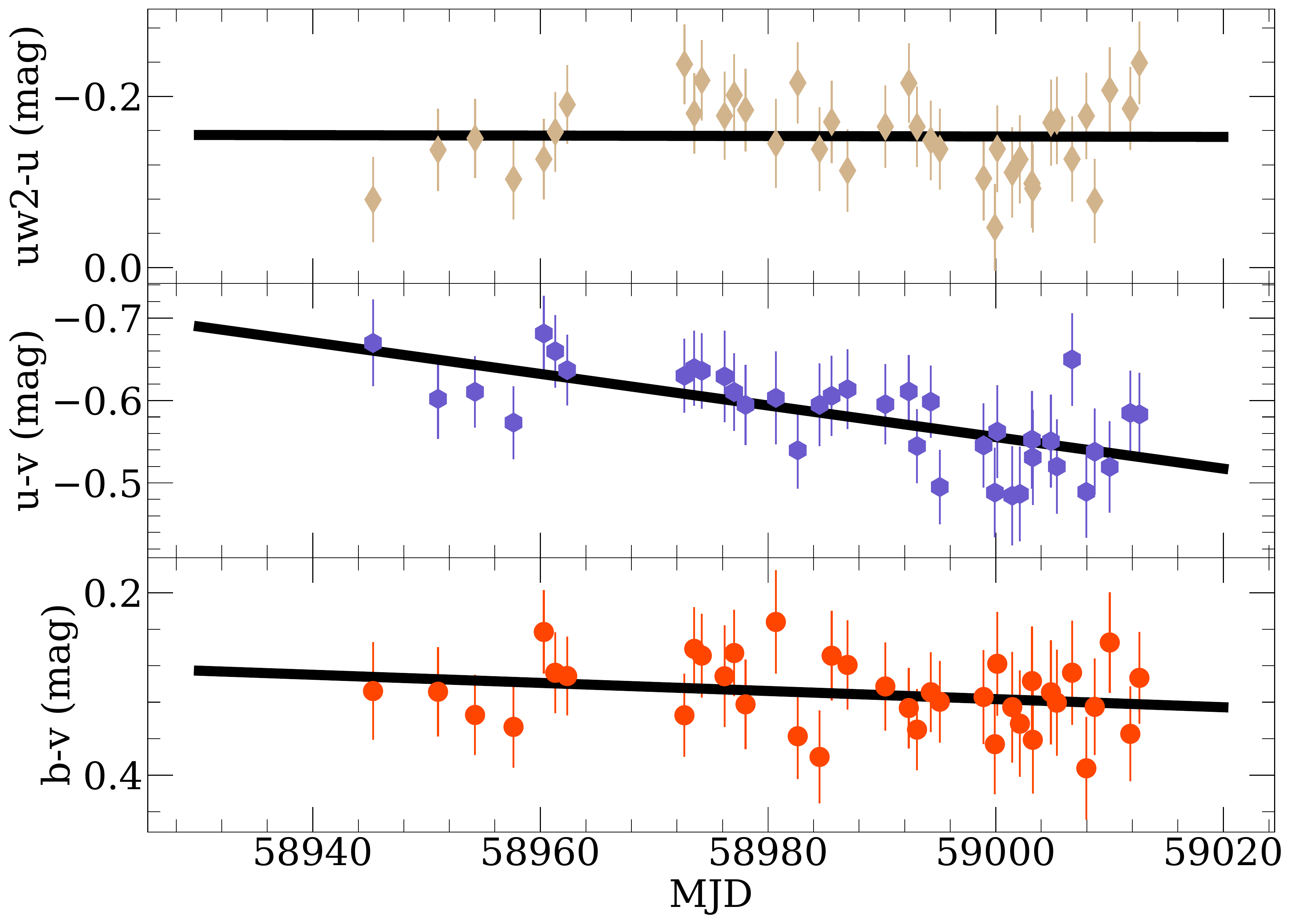}}
  \caption{colour evolution with time during the 2016--2017 and 2020 outburst. The black solid line in each panel denotes the linear fitting.}\label{fig:color_time}
\end{figure*}

The colours as a function of time are plotted in Figure~\ref{fig:color_time}. We noted that there is no significant ($\text{p}\gg 0.05$) variations of the colours with time in 2016--2017 outburst. However, in the case of 2020 outburst, a redden tendency can be seen in the case of \textsl{u-v} and \textsl{b-v}. Additionally, \textsl{u-v} shows a correlation ($\text{p}=4.051\times 10^{-7}$) with time, while \textsl{b-v} only manifests a weak (p=0.098) evolution. Similar to the trend of 2016--2017 outburst, there is no significant variations with time for \textsl{uw2-v}.

\section{Discussion}\label{sec:discussion}
The 2020 outburst has some common features with the 2016--2017 outburst. The 2020 outburst of OJ 287 has been studied previously and the after-effect of the black hole–disc impact was interpreted as the origin of the event \citep{komossa2020,komossa2021a,komossa2021b,kushwaha2020,prince2021b,prince2021a}. \citet{huang2021} suggested that the 2016--2017 outburst in OJ 287 was caused by a TDE. In this section, we compare the features of both the 2016--2017 and 2020 outbursts and combine these two outbursts to discuss the origin of the event as the after-effects of black hole–disc impact and the TDE. We find that the 2020 outburst in OJ 287 is probably caused by the TDE.

\subsection{Comparison with 2016--2017 outburst}
During 2016 October--2017 April, a sudden outburst covering from X-ray to radio bands in OJ 287 was observed \citep{komossa2017,kapanadze2018,kushwaha2018}. The 2016--2017 outburst expressed as high luminosity and ‘softer-when-brighter’ behaviour in X-rays \citep{komossa2017,komossa2020,kapanadze2018,kushwaha2018}. Additionally, during the outburst, unlike AGNs, the HR of OJ 287 exhibited a slight evolution with time and soft X-ray luminosity \citep{huang2021}. Furthermore, after the outburst, some emission lines in the optical spectra strengthened noticeably \citep{huang2021}.

There are some similarities between the 2016--2017 and 2020 outbursts, such as high luminosity, `softer-when-brighter' behavior in the X-rays and `bluer-when-brighter' in UV/optical bands. In addition, different from the case in the quiescent state, the correlation analysis of spectral index versus X-ray flux and HR versus soft X-ray flux of 2016--2017 and 2020 outbursts shows a similar behavior and reveals that both outbursts were dominated by the new component.

\begin{figure}
  \centering
  \includegraphics[width=0.45\textwidth]{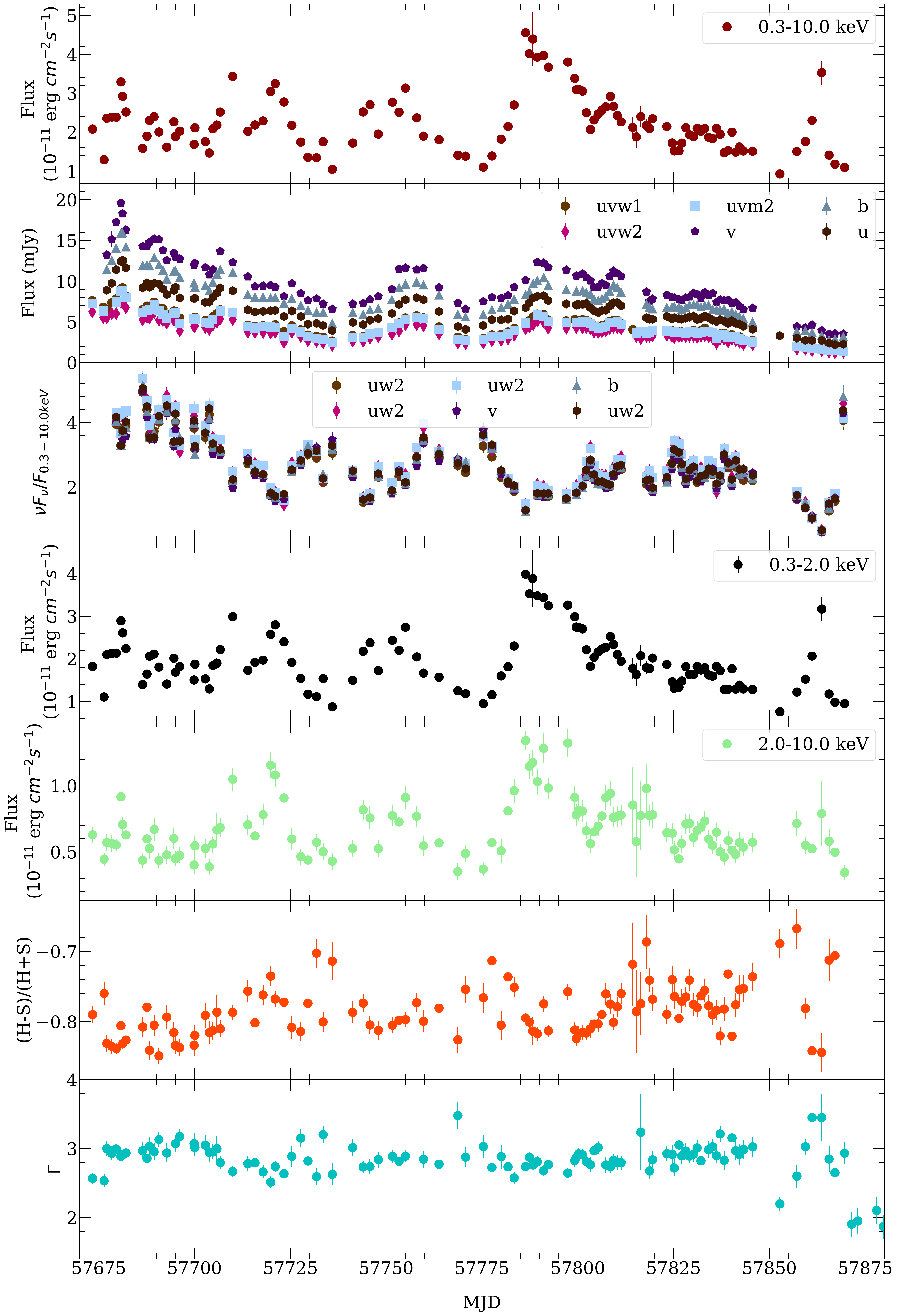}
  \caption{From top to bottom panels, the evolution of X-ray flux, UV/optical flux, ratio of UV/optical flux to X-ray, light curves of the soft and hard X-ray, HR, and spectral index during the 2016--2017 outburst.
  }\label{fig:2016flare}
\end{figure}

Figure~\ref{fig:2016flare} shows the evolution of the light curves, HR and spectral index in the 2016--2017 outburst. Compared with the 2020 outburst, the 2016--2017 outburst had a longer duration, a higher peak flux, a lower average HR, and a higher average value of the spectral index. In the early era of the 2016--2017 outburst, the ratio of UV/optical to X-ray flux was located at a high level and, after the time of the maximum X-ray flux, the flux ratio exhibited no significant variation. However, in the 2020 outburst, in the early era, the flux ratio of UV/optical to X-ray stated a constant level and, later in the epoch, it was greatly enhanced. In addition, during the second flare of the 2020 outburst, as the UV/optical flux increases, the X-ray spectral index decreases, and the HR increases, showing a different feature from the 2016–2017 outburst. During the 2016–2017 outburst, the UV/optical bands exhibited no significant evolution of colours while, in the case of the 2020 outburst, u-v and b-v redden with time. We integrated the flux, obtaining the energy released in the range of 0.3--10.0 keV as $1.386\times 10^{53}\text{erg}$ and $3.433\times 10^{52}\text{erg}$ for the 2016--2017 outburst and 2020 outburst, respectively. The summary of the comparison can be seen in Table \ref{tab:index_hr_evolution}.

\begin{table*}
\centering
\caption{The comparison of two outbursts, in 2016--2017 and 2020. The results are measured in the X-ray band (0.3--10.0 keV). The energy is integrated from MJD 57673 to MJD 57869 for the 2016–2017 outburst; for the 2020 outburst, we integrated from MJD 58945 to MJD 59012. Because the object is close to the Sun during the summer, there is no continuous monitoring, so we only present the lower limit of the duration of the outburst.}
\label{tab:comparison}
\begin{tabular}{lcc}
\hline
 &  2016--2017 & 2020 \\ \hline
Peak flux ( $10^{-11}\text{erg}~\text{cm}^{-2}~\text{s}^{-1}$)  & $4.553\pm{0.078}$ & $2.787\pm{0.074}$\\
Mean HR & $-0.785\pm{0.002}$ & $-0.754\pm{0.004}$ \\
Mean spectral index & $2.875\pm{0.016}$ & $2.510\pm{0.026}$ \\
Duration ($\text{days}$) & $196$ & $67$ \\
Energy ($\text{erg}$) & $1.386\times 10^{53}$ & $3.433\times 10^{52}$ \\ \hline
\end{tabular}
\end{table*}

\subsection{The aftereffect of black hole-disk impaction?}
OJ 287 was considered as an SMBHB system hosting an AGN \citep{sillanpaa1988}. In this system, the accretion disc of the primary black hole is impacted by the secondary twice in one period, and therefore the 12-yr quasi-periodic double-peak structure in the optical light curves can be observed \citep{lehto1996}. The observed optical outbursts in 2007, 2015, and 2019 coincide well with the theoretical predictions based on general relativity \citep{valtonen2008,valtonen2016,laine2020}.

When the secondary black hole punches into the disc, the perturbation caused by the event will propagate from the impacted site to the jet base, and thus the outburst triggered by the jet activity can be observed. This scenario was proposed to explain the 2016–2017 and 2020 outbursts in OJ 287 \citep{valtonen2017,kapanadze2018,kushwaha2018,kushwaha2020,komossa2020,komossa2021a,komossa2021b,prince2021b,prince2021a}.

\begin{figure}
  \centering
  \includegraphics[width=0.45\textwidth]{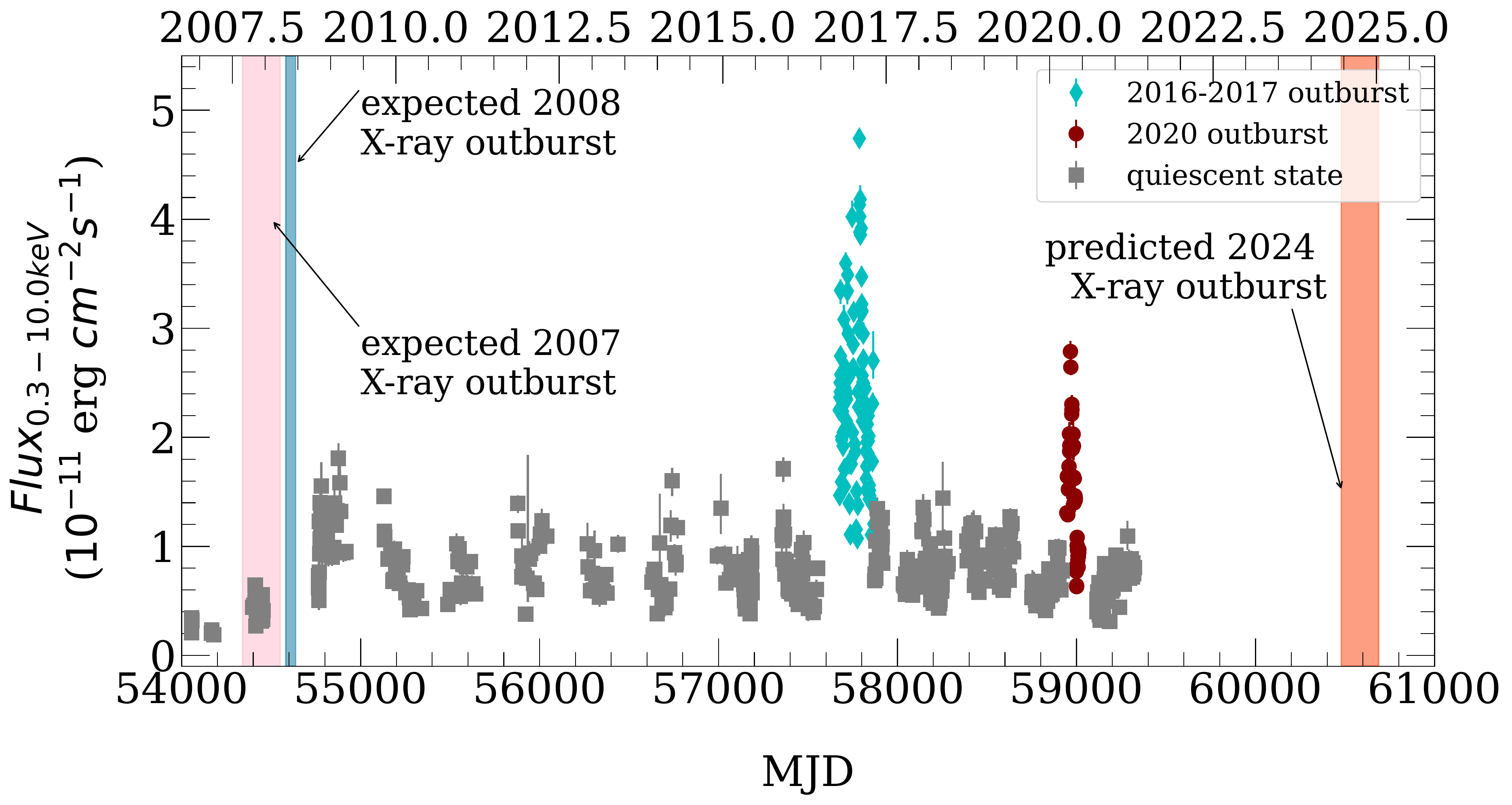}
  \caption{The predicted X-ray outburst induced by the after-effect of the black hole-disc impact. The colour shaded regions denote the range of the peak time. In the assumption, the expected X-ray outburst in 2007 is related to the impact in 2005, the expected X-ray outburst in 2008 is associated with the impact in 2007, and the 2022 impact will cause an X-ray outburst in 2024.}\label{fig:burst_predicted}
\end{figure}

If we assume that both the 2016--2017 and 2020 outbursts were related to the after-effect of the black hole impact, the velocity of the perturbation propagating across the disc can be roughly estimated. According to \citet{dey2018}, the optical outburst in 2015 December is related to the black hole–disc impact in 2013, and then another optical outburst that occurred in 2019 July is caused by the impact in that year. If the after-effect of the impacts in 2013 and 2019 induce the X-ray outbursts in 2016–2017 and 2020, then the average propagating velocity can be estimated. The average velocity of the perturbation can be calculated using $v=D/\Delta t$, where D is the distance of the impacted site from the primary black hole, and $\Delta t$ is the interval from the peak time of X-ray outburst to the impacted time. The distance from the primary black hole for the 2013 and 2019 impacts are 17566 and 3218 au, respectively \citep{dey2018}. Thus, we can obtain the propagation velocity of the perturbation as $v_{2013}=0.082c$ and $v_{2019}=0.066c$, and these two results can be considered as the upper and lower limits of the velocity. Therefore, we can obtain the peak time intervals of the X-rays that are related to the after-effect of the black hole–disc impacts. In Figure~\ref{fig:burst_predicted}, we show the predicted peak time of the X-ray outbursts.

The optical outburst in 2005 September was caused by the black hole–disc impact in 2005 March, and the expected X-ray outburst related to this event should peak at MJD 54340–54550 (pink region in Figure~\ref{fig:burst_predicted}). However, the observations show that the X-ray light curve is at a quiescent state during this time. Following the 2007 impact, it was expected that the after-effect would induced X-rays to peak at MJD 54581--54637; unfortunately, there are no data in this interval. In general, the perturbation undergoes energy dissipation as it propagates, so the loss of energy should be related to the distance of the propagation. The impact site in 2013 is further than the events in 2005, 2007, and 2019, and for this reason, there should be more energy dissipation as the perturbation crosses the disc to the base of the jet after the 2013 impact event. In other words, as the impact site is closer to the primary black hole, the less energy during propagation dissipates, the higher the impact velocity of the secondary black hole obtained, and the more energy the perturbation carries into the jet base -- thus, we should observe higher luminosity. The 2019 impact occurred at a much closer site than in 2013, but the 2020 outburst releases less energy than the 2016–2017 outburst. In addition, the 2016--2017 outburst lasted longer and had a higher peak luminosity. If the 2016--2017 and 2020 outbursts were caused by the after-effect of the 2013 and 2019 impacts, respectively, we should observe a much higher luminosity and more energy released in 2020. However, the observations are not consistent with this expectation. In addition, the periodic black hole–disc impact causes the 12-yr quasi-periodic optical outburst, and thus the outburst related to its after-effect should be periodic. However, only two huge soft X-ray outbursts are observed, which may not be a periodic event.

OJ 287 was a faint X-ray source in monitoring history until 2016 \citep{kapanadze2018}. \citet{sundelius1997} simulated the outbursts tidally induced by the pericentre passages of the secondary black hole in OJ 287. They predicted an outburst in late 2014 and early 2020, respectively. Additionally, \citet{sundelius1997} predicted that the outburst in late 2014 would have shorter duration and weaker luminosity than the outburst in early 2020. The observations of the 2016--2017 and 2020 outbursts are inconsistent with the results of the numerical predictions.

Nevertheless, the connection between the jet and the disc is still unclear, so we cannot estimate the peak time accurately. Within the after-effect of the black hole–disc impact, the predicted next peak time is MJD 60478--60688; if it is, then we will witness the next X-ray outburst at that time. However, the inconsistency between previous observations and predictions suggests that the 2016--2017 and 2020 outbursts may not be related to the after-effect of the black hole-disc impact. Therefore, we should not expect so much for the next huge X-ray outburst in the predicted interval.

\subsection{TDE in OJ 287?}
Considering a star with mass $M_*=m_* M_{\sun}$ and radius $R_*=r_* R_{\sun}$, which is torn by a supermassive black hole with the mass $M_{BH}$. The minimum fallback time-scale of the stellar debris in a TDE can be written as
\begin{equation}\label{eq:mini_fallback}
T_{min}=41M_6^{\frac{1}{2}}m_{*}^{-1}\beta^{-3}r_{*}^{\frac{3}{2}} ~\text{days},
\end{equation}
where $M_6=\frac{M_{BH}}{10^6M_{\sun}}$, $\beta=\frac{r_t}{r_p}$ is the penetration factor, $r_t=\left(\frac{M_{BH}}{M_*}\right)^{\frac{1}{3}}R_*$ is the tidal radius and $r_p$ is the pericentre of the star \citep{lodato2011}. Defining $t=\frac{T-T_0}{T_{min}}$, where $T_0$ is the disrupted time, we focus on the decay of the light curves evolving with $t$.

\subsubsection{Light curve fitting by TDE prediction}
It is hard to interpret the 2016--2017 outburst as one of the common events in OJ 287, because such a phenomenon has never before been observed in this object. \citet{huang2021} suggested that the 2016--2017 outburst was caused by a TDE in the supermassive black hole system of OJ 287. During the outburst, the X-rays exhibited low HR values and steep X-ray spectra. The X-ray spectra in the outburst can be fitted well by the double power-law model, and the second component exhibits extremely soft X-rays, suggesting that the new component is different from previous observations of this object. Furthermore, the negligible evolution of the HR with soft X-ray luminosity, the prominent strengthening of emission lines in optical spectra after the outburst, and the fact that the decay of the light curves is well fitted by
$t^{-5/3}$ to the decay of the light curves, suggest that the 2016--2017 outburst was probably caused by a TDE.

\begin{figure*}
  \centering
  \subfigure [2016--2017 outburst] {
  \includegraphics[width=0.45\textwidth]{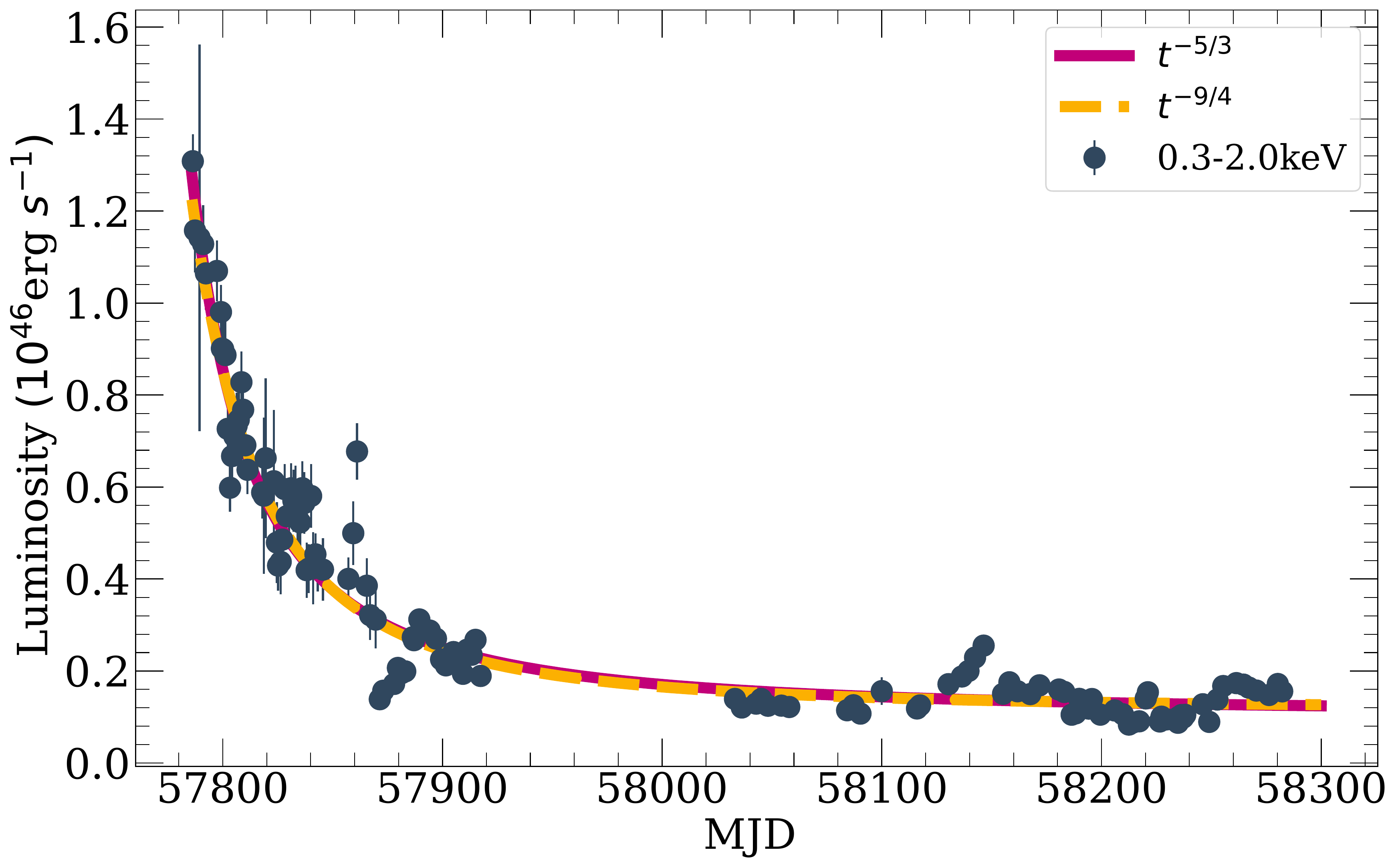}}
  \subfigure [2020 outburst] {
  \includegraphics[width=0.45\textwidth]{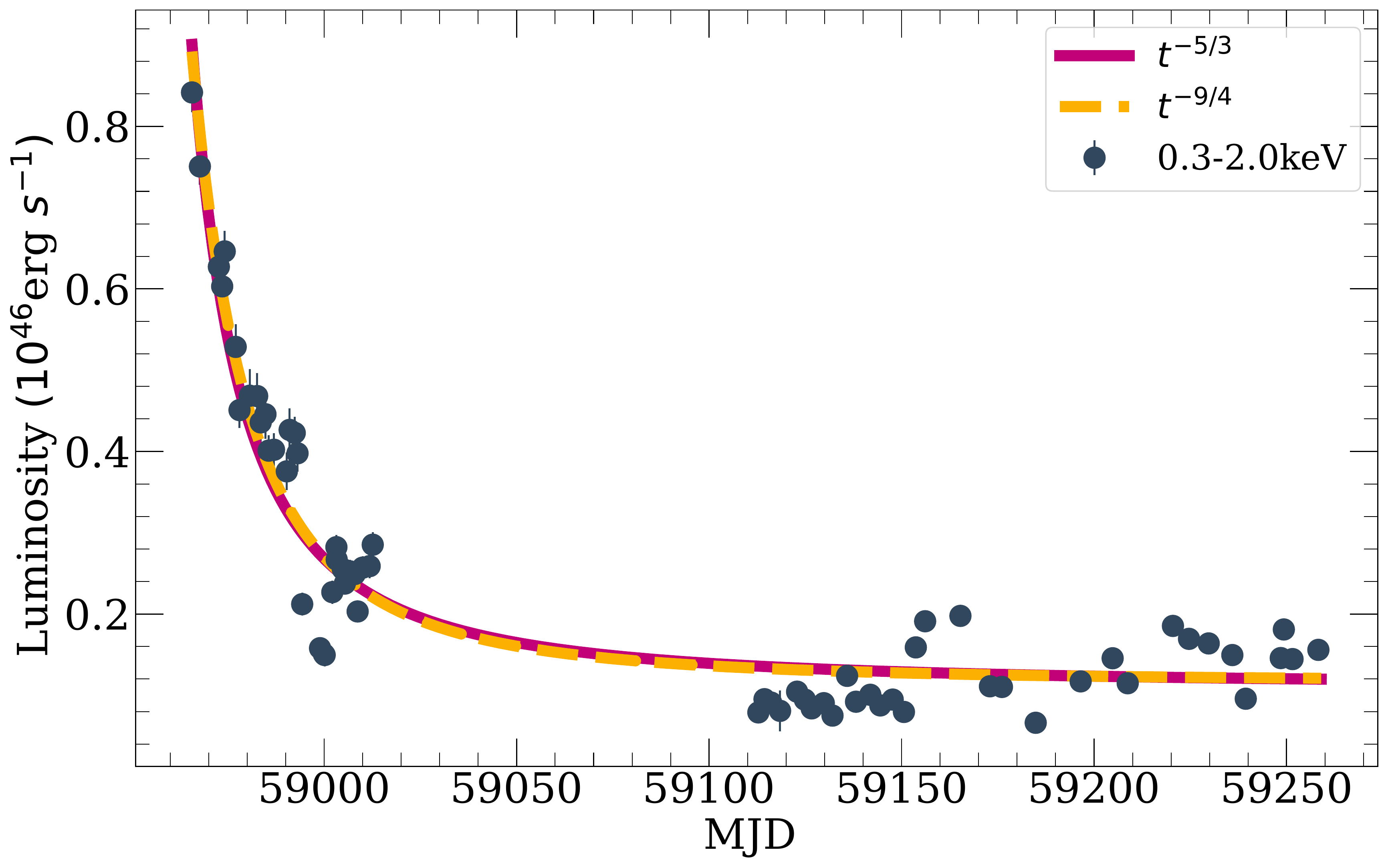}}
  \caption{The fitted results for the decay portion of the soft X-ray light curves. The light curves are fitted by both $t^{-5/3}$ and $t^{-9/4}$. The fitted results for the 2016--2017 and 2020 outburst are plotted in panel (a) and panel (b), respectively.}\label{fig:x_fit}
\end{figure*}

The mass of the primary black hole in OJ 287 was calculated as $1.83\times 10^{10} M_{\sun}$, and the secondary as $1.5 \times 10^{8} M_{\sun}$ \citep{dey2018}. \citet{chen2021} suggest that for the black hole with the mass $ > 10^8 M_{\sun}$, only partial tidal disruption event (PTDEs) can occur.
For this reason, the remnant of the disrupted star may survive after TDE occurred in OJ 287. Therefore, as a result of orbiting around the black hole, the remnant would be disrupted again when it approaches the tidal radius. Recently, five quasi-periodic eruptions (QPEs) have been found: GSN 069 \citep{miniutti2019}, RXJ1301.9+2747 \citep{giustini2020}, eRO-QPE1, eRO-QPE2 \citep{arcodia2021} and XMMSL1 J024916.6-041244 \citep{chakraborty2021}. PTDEs are considered as their origin \citep{zhao2021}, as in ASASSN-14ko \citep{payne2021}. The 2016--2017 outburst may probably be caused by a PTDE, and thus the 2020 outburst may have the same event because it has common features. Different from the full TDE, the light curve of PTDE decays following $t^{-9/4}$ \citep{coughlin2019}. We fitted the decay light curves of the soft X-rays for the outbursts in 2016--2017 and 2020 with $t^{-5/3}$ and $t^{-9/4}$, respectively. For the 2016--2017 outburst, the coefficients of determination ($R^2$) in the fitting are 0.943 and 0.944, fitted by $t^{-5/3}$ and $t^{-9/4}$, respectively. For the 2020 outburst, we obtained the values of $R^2$ as 0.915 and 0.921 for the fitted results by $t^{-5/3}$ and $t^{-9/4}$, respectively. There is no significant difference for $R^2$ when the light curves fitted by $t^{-5/3}$ and $t^{-9/4}$. The fitted results are shown in Figure~\ref{fig:x_fit}.

\subsubsection{The progenitor star}
We integrated the luminosity during the outburst and obtained the released energy in 0.3--10.0 keV of the value as $1.386\times 10^{53}$ erg and $3.433\times 10^{52}$ erg, for the 2016--2017 and 2020 outburst, respectively. Through $\Delta E=\eta \Delta m c^2$, we can estimate the accreted mass from the disrupted star.
The mean flux ratio of v band to X-ray (0.3--10.0 keV) are 2.546 and 3.224, for 2016--2017 and 2020 outburst, respectively. Thus the contribution fraction of the X-ray in the released energy in two outbursts are 0.282 and 0.237, respectively. Assuming $\eta=0.1$, through the energy released in the X-ray band (0.3--10.0 keV), we estimated that the accreted masses are $2.74M_{\sun}$ and $0.81M_{\sun}$ in 2016--2017 and 2020 outbursts, respectively.

The pericentre of the star can be written as
\begin{equation}
r_p=\left(\frac{GM_{BH}P^2}{4\pi^2}\right)^{\frac{1}{3}}(1-e),
\end{equation}
where $G$ is the gravitational constant, $P$ is the the period of the star, and $e$ is the eccentricity. Then, the penetration factor can be rewritten as
\begin{equation}\label{eq:penetration}
\beta=\frac{(4\pi^2)^{\frac{1}{3}}R_{\sun} r_*}{G^{\frac{1}{3}}P^{\frac{2}{3}}
M_{\sun}^{\frac{1}{3}}m_{*}^{\frac{1}{3}}(1-e)}.
\end{equation}
From equation (\ref{eq:mini_fallback}), penetration factor also can be obtained as
\begin{equation}\label{eq:penetration2}
\beta=\frac{41^{\frac{1}{3}}M_6^{\frac{1}{6}}m_{*}^{-\frac{1}{3}}
r_{*}^{\frac{1}{2}}}{T_{min}^{\frac{1}{3}}}.
\end{equation}
From both equations (\ref{eq:penetration}) and  (\ref{eq:penetration2}), the expression of the stellar radius can be written as
\begin{equation}\label{eq:stellar_radius}
r_*=\frac{41^{\frac{2}{3}}M_6^{\frac{1}{3}}G^{\frac{2}{3}}
P^{\frac{4}{3}}M_{\sun}^{\frac{2}{3}}(1-e)^2}
{T_{min}^{\frac{2}{3}}(4\pi^2)^{\frac{2}{3}}R_{\sun}^2}.
\end{equation}

\begin{figure}
  \centering
  \includegraphics[width=0.45\textwidth]{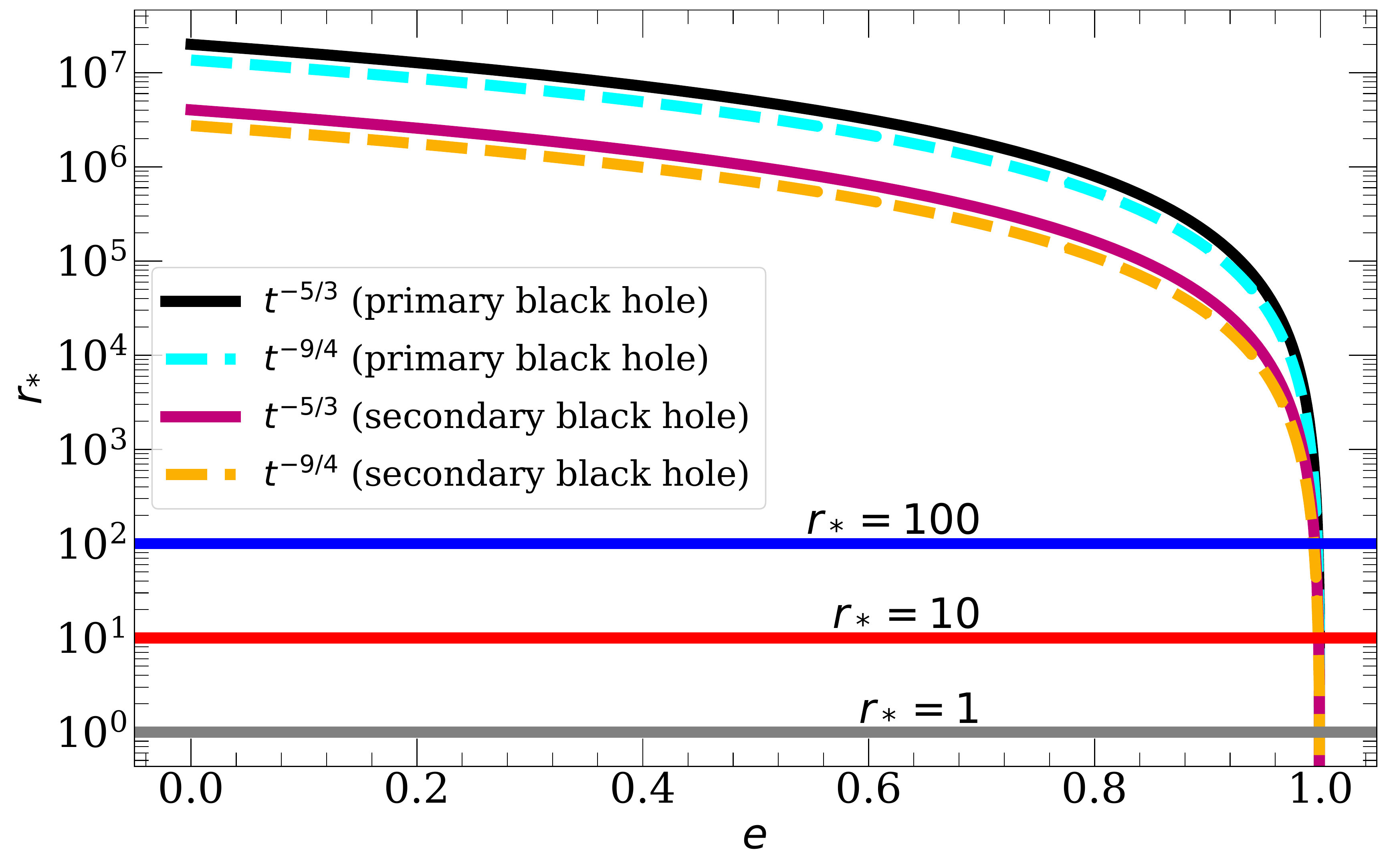}
  \caption{The stellar radius as a function of eccentricity. The results are plotted from equation (\ref{eq:stellar_radius}), while the values of $T_{min}$ are extracted from the light curve fitting by $t^{-5/3}$ and $t^{-9/4}$ for both cases of the primary and secondary black hole. The three solid lines denote $r_*=100$ (blue), $r_*=10$ (red), and $r_*=1$ (grey), respectively.}\label{fig:r_e2020}
\end{figure}

Assuming that the star passes the pericentre twice at the same distance, the period was set as the interval of the maximum luminosity time in optical bands between the 2016--2017 and 2020 outburst, and we obtained $P$ of a value of 1281 d (the maximum luminosity times of optical bands are MJD 58962 and MJD 57681, respectively). Additionally, with the mass of the primary and secondary black holes having values of $1.83\times10^{10}M_{\sun}$ and $1.5\times 10^8 M_{\sun}$, respectively, the relation between the eccentricity and stellar radius is shown in Figure~\ref{fig:r_e2020}. We can see that the stellar radius depends on the eccentricity of its orbit. Denser stars are allowed in the TDE when they are in a high eccentric orbit.  Within three typical values of stellar radius, $r_*=1$, $r_*=10$ and $r_*=100$, we obtain the eccentricity of 0.9998, 0.9993, and 0.9978 for the primary black hole, and the values of 0.9995, 0.9984, and 0.9950 for the secondary black hole, respectively. In the case of OJ 287, to ensure the stellar radius is in a reasonable range, high eccentricity is needed. These results are consistent with \cite{Bogdanovic2014}.

\cite{macLeod2013} predicted that in every passage through pericentre, the mass loss of the star is
\begin{eqnarray}\label{eq:mass_loss}
\Delta M(\beta)=f(\beta)\left(\frac{M_{*}-M_{c}}{M_*}\right)^2M_{*},
\end{eqnarray}
where $M_c$ is the mass of the stellar core, and
\begin{eqnarray}
f(\beta)=\left\{
  \begin{array}{ll}
    0, & \beta < 0.5, \\
    \frac{\beta}{2}-\frac{1}{4}, &  0.5 \leq \beta \leq 2.5, \\
    1, &  \beta > 2.5.\\
  \end{array}
\right.
\end{eqnarray}

Substituting equation (\ref{eq:penetration}) into equation (\ref{eq:mass_loss}), we obtain the stellar radius in another form as
\begin{equation}\label{eq:radius_e_core}
r_*=\frac{(1-e)G^{\frac{1}{3}}P^{\frac{2}{3}}\left[4\Delta M+m_*M_{\sun}\left(1-\frac{m_c}{m_*}\right)^2\right]}{2m_*^{\frac{2}{3}}
M_{\sun}^{\frac{2}{3}}(2\pi^2)^{\frac{2}{3}}\left(1-\frac{m_c}{m_*}\right)^2R_{\sun}},
\end{equation}
where $m_c=\frac{M_c}{M_{\sun}}$.

\begin{figure*}
  \centering
  \subfigure [$e=0.995$] {
  \includegraphics[width=0.45\textwidth]{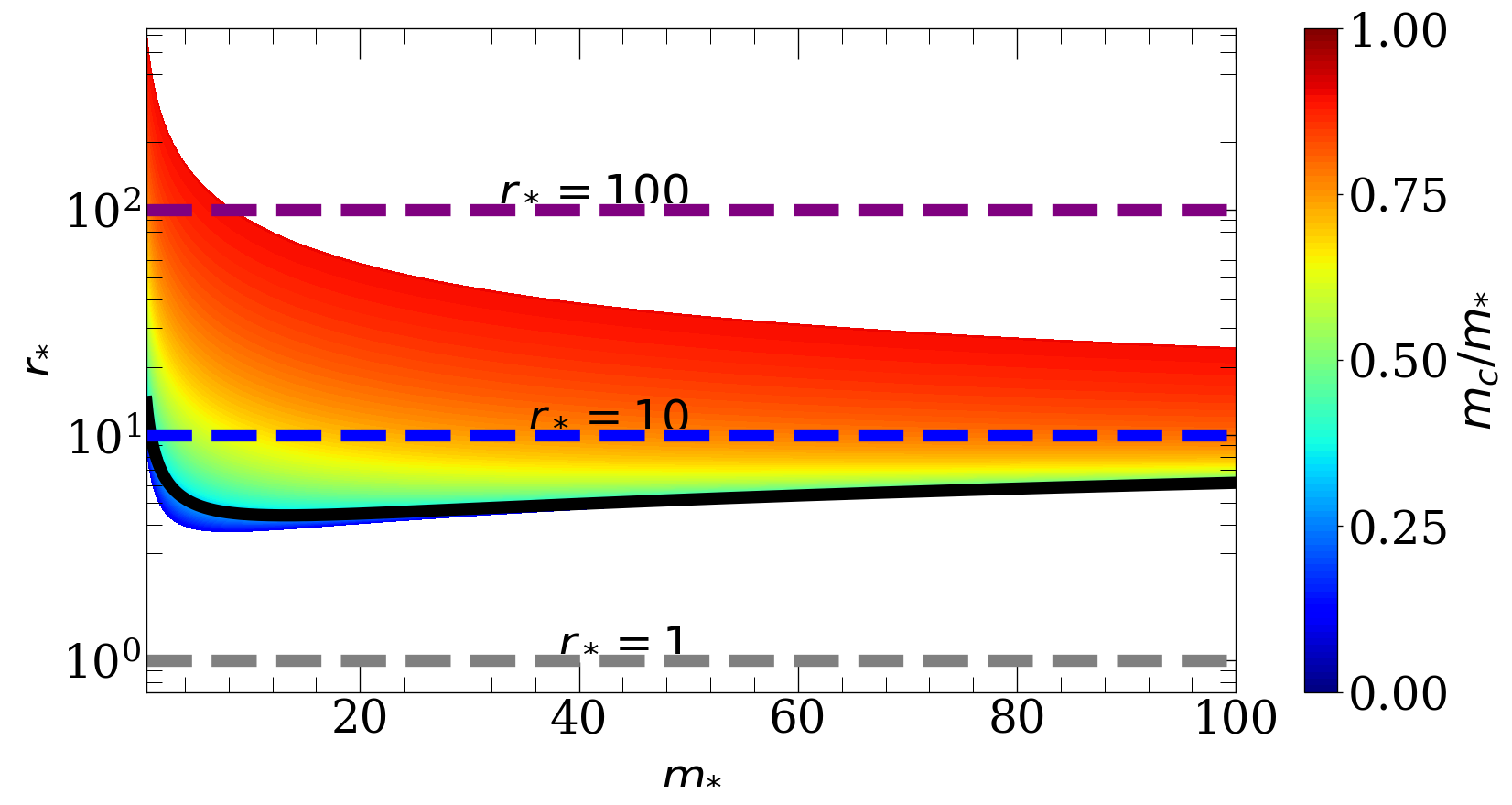}}
  \subfigure [$e=0.9995$] {
  \includegraphics[width=0.45\textwidth]{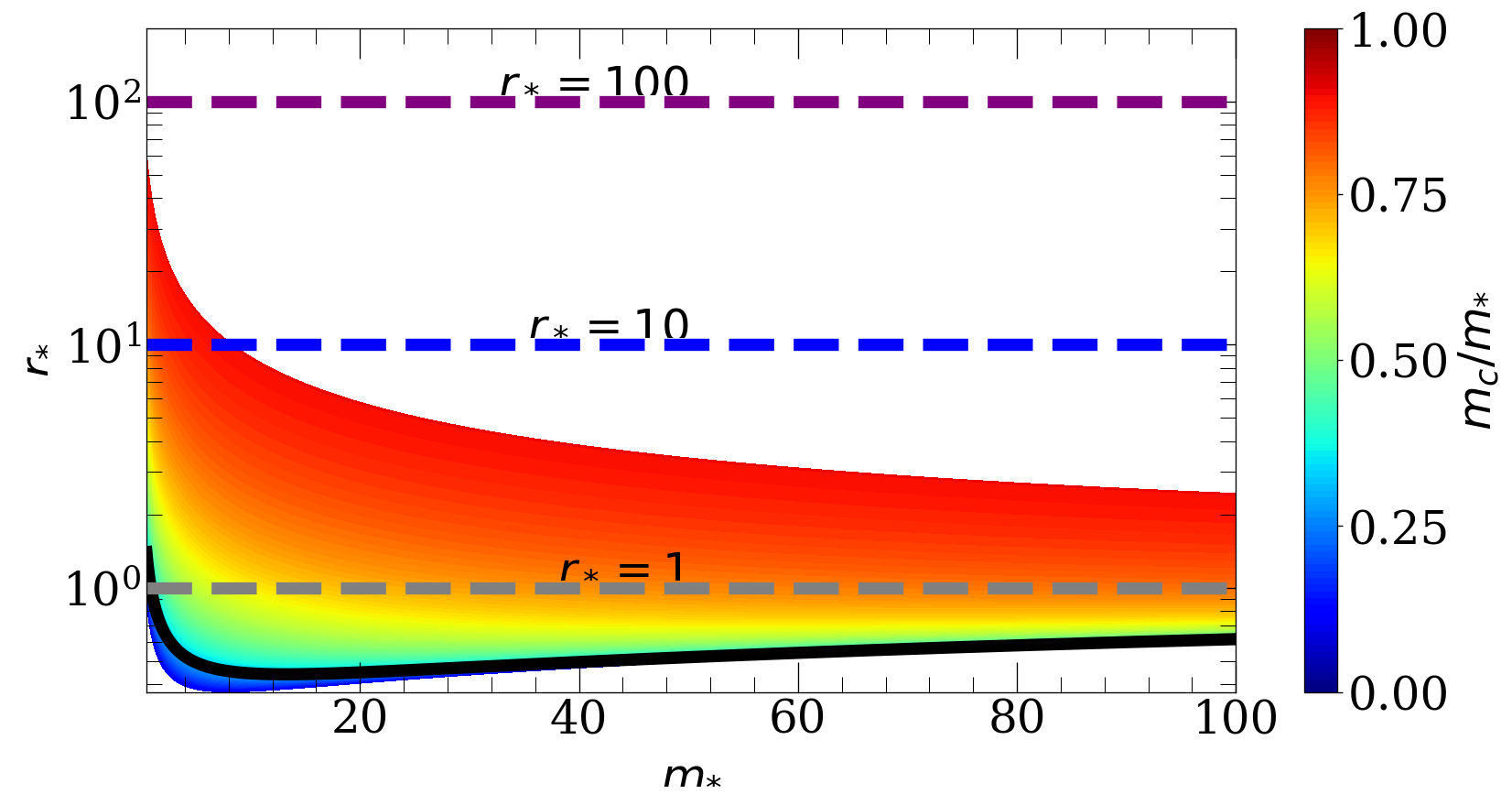}}

  \subfigure [$\frac{m_c}{m_*}=0.1$] {
  \includegraphics[width=0.45\textwidth]{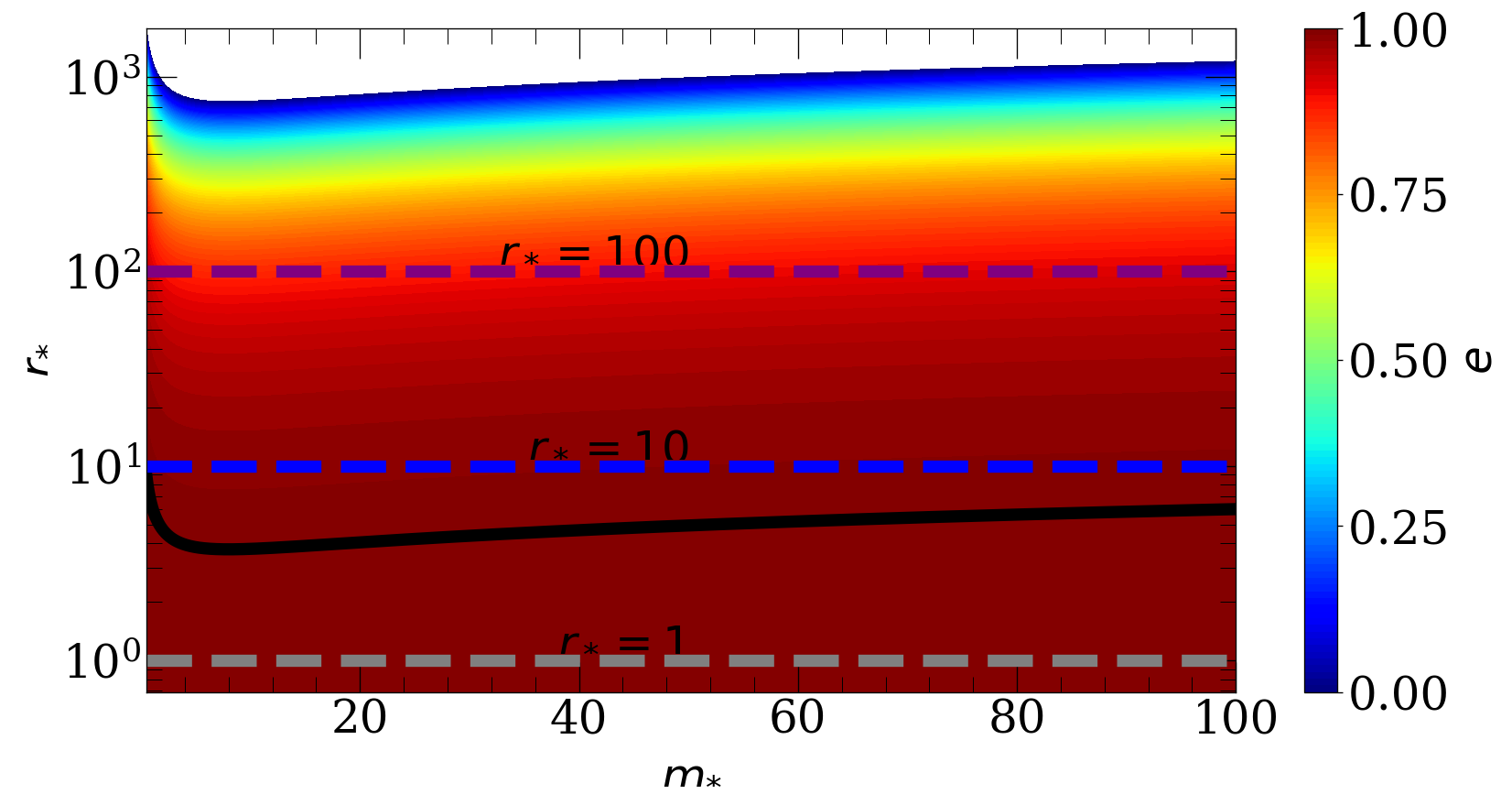}}
  \subfigure [$\frac{m_c}{m_*}=0.3$] {
  \includegraphics[width=0.45\textwidth]{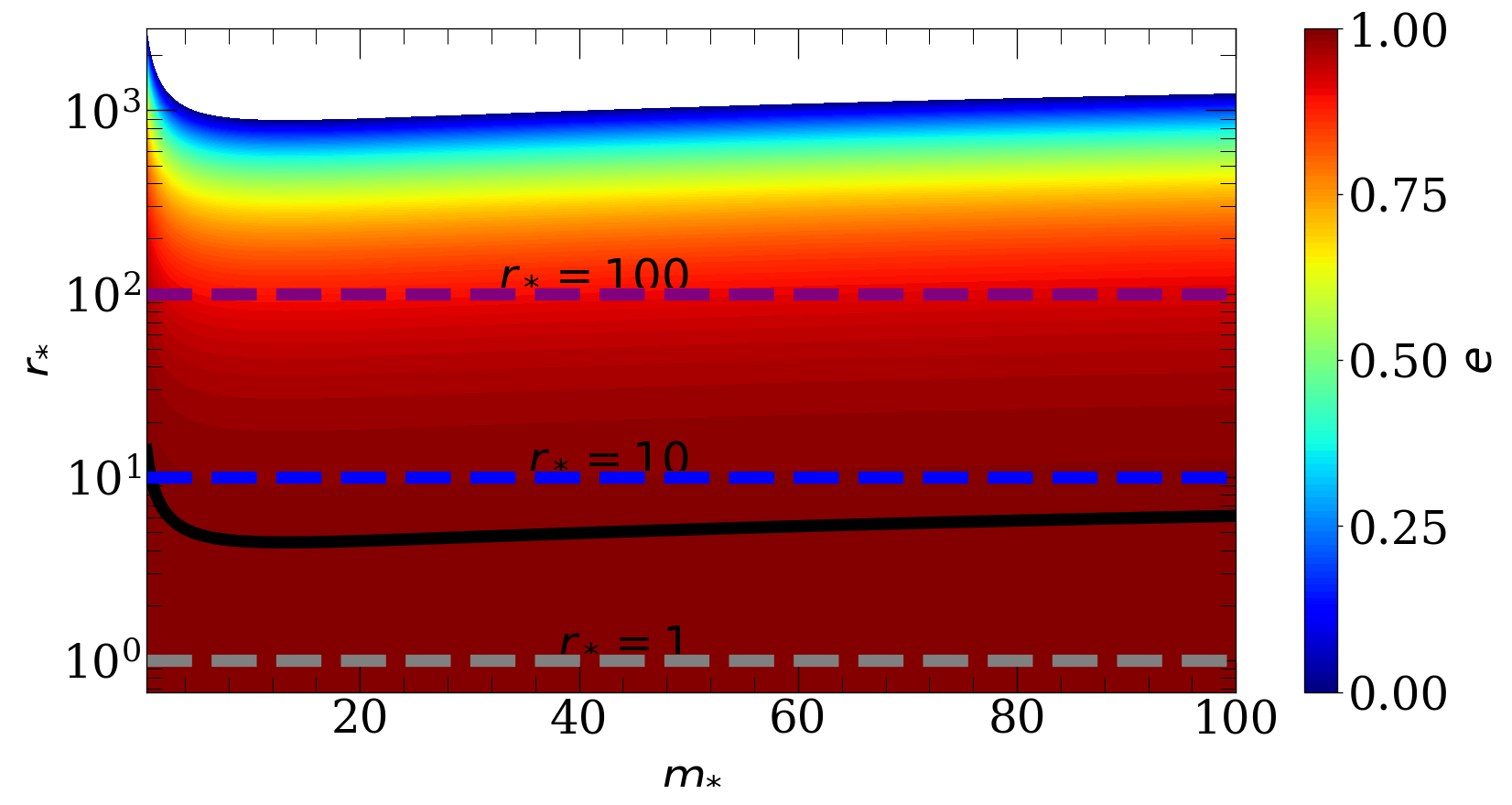}}
  \caption{The relation between radius and mass of the disrupted star, within different eccentricity and core mass. In panels (a) and (b), fixing the eccentricity as the values of 0.9950 and 0.9995, respectively, the relation between stellar radius, mass and core mass are plotted. The black solid curves in panels (a) and (b) plot the relation with $\frac{m_c}{m_*}=0.3$.
In panels (c) and (d), the values of $\frac{m_c}{m_*}$ are fixed as 0.1 and 0.3, then the relations between radius, mass and eccentricity are plotted. The black curves in panels (c) and (d) plot the evolution of radius with mass fixing the eccentricity as 0.995. The dashed lines in each panel denote the $r_*=100$ (blue), $r_*=10$ (purple) and $r_*=1$ (grey), respectively.}\label{fig:radius_e_core}
\end{figure*}

In equation (\ref{eq:radius_e_core}), the stellar radius is affected by its mass, core mass, eccentricity, period of the star and the accreted mass in each pericentre passage. With the accreted mass of $0.81M_{\sun}$, we plot the evolution of stellar radius as the function of stellar mass with variation of core mass and eccentricity in Figure~\ref{fig:radius_e_core}.
At a certain eccentricity and $m_c<0.7$, the radius of stars decreases with mass, while for high-mass stars, the radius increases slowly with mass.
However, for the case of $m_c>0.7$, the radius declines with the stellar mass. The allowed stellar radius increases with the fraction of the core mass. Additionally, the relations between stellar radius and mass depends on eccentricity. If we assume that the pericentre equals the Schwarzschild radius, the eccentricity can be estimated as 0.9976. Therefore, if the TDE interpretation is correct, the disrupted star may be of several times the solar mass, and its orbital eccentricity may be high.

\subsubsection{Some issues in the TDE hypothesis}
There are some issues with this scenario. First, if the 2020 outburst is related to a PTDE, the estimated accreted mass may be too large. In the first disruption, the envelope of the star is torn and accreted into the black hole, and then the surviving remnant would be denser and hard to be disrupted again. It is also possible that in the first disruption, part of the envelope material is incompletely disrupted, and the remaining part is destroyed again the second time.

Second, the observation in the radio band \citep{prince2021a} showed variability in radio of 15-GHz during the 2020 outburst and this may be the signature of jet activity. Additionally, \citet{komossa2020} reported the Fe absorption line nearing the peak luminosity of the X-ray, suggesting an outflow with a velocity of $\sim 0.1c$. However, the existence of an outflow was found in some TDE candidates, such as ASASSN-14li \citep{velzen2016}, ASASSN-15oi \citep{horesh2021}, and AT2019dsg \citep{stein2021}, observed in radio bands. Thus, we cannot completely rule out the possibility that the signature of an outflow originates from the TDE.

The third issue is the rebrightening of the UV/optical bands in the later epoch of the light curves. During the decline phase of the X-ray, the UV/optical light curves rebrighten and there is significant enhancement in the HR. We do not see these features in the 2016--2017 outburst. The phenomenon of rebrightening in the UV/optical bands is common in TDE candidates,for example, ASASSN-15lh \citep{leloudas2016}, PS16dtm \citep{blanchard2017}, and AT2019avd \citep{malyali2021}. \citet{chen2021b} suggested the first peak is related to the circularization process, and the second peak is caused by the delay in accretion. The hardened trend of X-rays when the rebrightening occurred in UV/optical bands is related to the decrease of the fraction of the soft X-rays, while the hard X-ray state is constant. The hardening of X-rays in the late epoch has been found in 1ES 1927+654 and it is interpreted as the result of the reformation of the corona \citep{ricci2020}. However, in the case of OJ 287, the X-ray flux did not increase like the HR, and only the enhancement in the UV/optical bands was observed. This may be associated with the reprocessing of X-rays, which transforms the radiation into UV/optical bands.

The fourth is the colour evolution. At both outbursts (2016–2017 and 2020), the ‘bluer-when-brighter’ tendency is observed, and this phenomenon may be caused by the variation in the accretion rate \citep{pereyra2006,hung2016}. In the 2016--2017 outburst, there is no significant temporal evolution of colours, which is consistent with the scenario of TDEs \citep{blanchard2017,zabludoff2021}. However, in the 2020 outburst, the reddening of \textsl{u-v} and \textsl{b-v} with time may suggest a different mechanism compared with the 2016–2017 outburst. Considering the rebrightening of the UV/optical flux and the sign of the outflow in the X-ray spectrum, the reddening may be related to the additional activity of the jet or the interaction between the outflow and the environment of the blazar. Thus, the temporal reddening behaviour of \textsl{u-v} and \textsl{b-v} does not conflict with the TDE hypothesis.

Fifthly, the existing disc would block the emission from the TDE by the secondary black hole. If the TDE hypothesis holds, then the TDE events that trigger the 2016--2017 and 2020 outbursts occur on either side of the accretion disc. After the black hole impact in 2019, the secondary black hole crossed the disc to its other side. If the TDE occurred at that time, then the radiation from the disruption may be obscured by the disc and it would be difficult to observe the event. One of the possibilities is that the star may be disrupted by the primary black hole and, as a result, without being obscured, the radiation in the TDE can be detected directly. With the primary black hole spin of 0.33 \citep{dey2018}, we estimated the ratio of the tidal radius to the innermost stable circular orbit (ISCO) of 1, while $m_*=10$ and $r_*=160$. This suggests that the radius needs to be large enough to be disrupted by the primary black hole rather than swallowed. The last black hole impact occurred in the middle of 2019, with the distance from the impact site to the primary black hole of 3218 au ($18 r_g$, where $r_g$ is the gravitational radius); if the 2020 outburst is related to the TDE, the orbiting remnant may be influenced by the primary black hole. Therefore, if the radius of the remnant is large, it may be disrupted by the primary black hole at that time. To ensure that disruption rather than swallowing occurred in the primary black hole, the density of the material should be low enough. Another possibility is that the remnant is disrupted by the secondary black hole. The location of the TDE is in the vicinity of the primary black hole, so the debris stream is affected by its gravity and falls back to the primary one \citep{liu2009,coughlin2017,coughlin2019b}. For this reason, we can observe radiation in the TDE even if the secondary is obscured by the disc of the primary black hole.

Although most of the discussions involving the TDE are based on main-sequence stars, authors also found TDEs related to giant stars \citep[e.g. PS1-10jh][]{Bogdanovic2014}. The rate for a giant star disrupted by a black hole is higher than a main-sequence star \citep{stone2020}. Additionally, the TDE rate in AGNs is much higher than that in normal galaxies \citep{kennedy2016}. Despite the fact that there are some features different from common TDEs, we can still consider the TDE hypothesis as the probable explanation for the 2020 outburst.

\section{Conclusion}\label{sec:conclusion}
We investigated the 2020 outburst in OJ 287 in this work. Through the multiwavelength light curves, X-ray spectral index and the HR, we explore the origin of this outburst. Two scenarios, including the after-effect of black hole–disc impact and the TDE, are discussed as possible candidates.

The outburst mainly covered from the X-ray to radio bands while there is no significant variation in $\gamma$-rays. The X-rays exhibit a ‘softer-when-brighter’ behaviour and this phenomenon has been reported in \citep{komossa2020,komossa2021a,komossa2021b,kushwaha2020}. The rebrightening of the UV/optical light curves was detected during the decreasing X-ray phase. In the rebrightening of the UV/optical bands, the flux ratio of UV/optical to X-ray rises and the X-rays show a tendency to harden at that time. The correlation analyses of the spectral index versus X-ray flux and the HR versus soft X-ray flux reveal that the 2020 outburst was dominated by a new component, as in the case of the 2016–2017 outburst. The UV/optical bands show the ‘bluer-when-brighter’ tendency and this feature has been reported by \citet{prince2021b}. The 2020 outburst manifests a reddening with time in \textsl{u-v} and \textsl{b-v}, while in the 2016–2017 outburst, no significant evolution of colours has been seen.

The black hole–disc impact model has been successfully applied to explain the 12-yr quasi-periodic optical outburst with a double-peak structure. Furthermore, its after-effect is considered to be related to the 2016–2017 outburst \citep{komossa2017,kapanadze2018,kushwaha2018} and the 2020 outburst \citep{komossa2020,komossa2021a,komossa2021b,kushwaha2020,prince2021b,prince2021a}. According to the after-effect of the black hole–disc impact, after every impact, X-ray outbursts similar to the 2016–2017 and 2020 outbursts should be observed. However, historical observations do not show huge soft X-ray outbursts before 2016, so we argue that the outbursts in 2016–2017 and 2020 might not be caused by the after-effect of the black hole–disc impact.

The X-ray decay of the 2020 outburst can be fitted well by the prediction of full TDE ($t^{-5/3}$) and PTDE ($t^{-9/4}$). The low HR, steep X-ray spectra, and no significant HR evolving with soft X-ray flux are consistent with the X-ray properties of TDE candidates \citep{auchettl2017,auchettl2018}. However, the optical bands show the tendency to redden with time in the 2020 outburst, while TDE candidates manifest inconspicuous evolution of colours \citep{gezari2021,zabludoff2021}. The 2016--2017 outburst is explained well by the TDE scenario, from the HR, spectra and light curve evolution \citep{huang2021}. If the 2020 outburst was caused by the TDE, the disrupted material should be provided by the remnants from PTDE, which caused the 2016–2017 outburst. Therefore, a red giant star was probably to be disrupted and its orbital eccentricity should be high ($e >0.99$). Through the analysis, we find that the TDE might be the probable origin of the 2020 outburst in OJ 287.

We provide the TDE scenario to explain the 2020 outburst in OJ 287 and further monitoring and research are encouraged. As a well-known blazar, OJ 287 provides an excellent natural laboratory for studying the physical mechanisms of SMBHB theory, general relativity, jets and accretion discs. The study of the 2020 outburst contributes to the understanding of the origin of variability, which is a good reference for the study of the physical mechanisms of other AGNs.

\section*{Acknowledgements}
We are grateful to the anonymous referee for valuable
comments that improved the quality of this work. We thank Professors Rongfeng Shen and Xinwen Shu for the valuable advices. This work is supported by the Natural Science Foundation of China under grant No.~11873035 and the Natural Science Foundation of Shandong province (No.~JQ201702). S. Alexeeva acknowledges support from the National Natural Science Foundation of China (grant No.~12050410265) and LAMOST FELLOWSHIP program that is budgeted and administrated by Chinese Academy of Sciences.
We acknowledge Swift for the observation and providing the public data for the research.

%%%%%%%%%%%%%%%%%%%%%%%%%%%%%%%%%%%%%%%%%%%%%%%%%%
\section*{Data Availability}

The data analysed in this study can be freely retrieved from public archival database. The $\gamma$-ray data are downloaded from the official websites at \url{https://fermi.gsfc.nasa.gov/ssc}. The X-ray and UV/optical data are download from the website of HEASARC website at \url{https://heasarc.gsfc.nasa.gov/db-perl/W3Browse/w3browse.pl}. The python module {\it emcee} was taken from the website \url{https://emcee.readthedocs.io}.

\bibliographystyle{mnras}

\bibliography{oj287}
\end{document}